\documentclass[12pt]{article}%
\usepackage{amsmath}
\usepackage{amsfonts}
\usepackage{amssymb}
\usepackage{graphicx}%
\providecommand{\U}[1]{\protect\rule{.1in}{.1in}}
\newcommand{\bfr}{\begin{flushright}}
\newcommand{\efr}{\end{flushright}}
\newcommand{\bc}{\begin{center}}
\newcommand{\ec}{\end{center}}
\newcommand{\ben}{\begin{enumerate}}
\newcommand{\een}{\end{enumerate}}

\newcommand{\be}{\begin{equation}}
\newcommand{\ee}{\end{equation}}
\newcommand{\ba}{\begin{array}}
\newcommand{\ea}{\end{array}}
\def\6{\partial}
\begin{document}

\title{\textbf{Nonequilibrium dynamics of strings }\\\textbf{in time-dependent plane wave backgrounds}}
\author{R. Nardi$^{*}$ and I. V. Vancea$^{**}$\\\thanks{rnardi@cbpf.br, $^{**}$ionvancea@ufrrj.br} \emph{{\small Centro
Brasileiro de Pesquisas F\'isicas (CBPF) }} \\\emph{{\small R. Dr. Xavier Sigaud 150, 22290-180 Rio de Janeiro - RJ,
Brasil}}\\$^{**}$\emph{{\small Grupo de F{\'{\i}}sica Te\'{o}rica e Matem\'{a}tica
F\'{\i}sica, Departamento de F\'{\i}sica, }} \\\emph{{\small Universidade Federal Rural do Rio de Janeiro (UFRRJ),}} \\\emph{{\small Cx. Postal 23851, BR 465 Km 7, 23890-000 Serop\'{e}dica - RJ,
Brasil }} }
\date{19 December 2011}
\maketitle

\thispagestyle{empty}
%\pagestyle{empty}

%\begin{center}
%\rule{15cm}{0.01cm}
%\end{center}

\abstract{We formulate and study the nonequilibrium dynamics of strings near the singularity of
the time-dependent plane wave background in the framework of the Nonequilibrium Thermo Field Dynamics (NETFD).
In particular, we construct the Hilbert space of the thermal string oscillators at nonequilibrium and generalize the
NETFD to describe the coordinates of the center of mass of the thermal string. The equations of
motion of the thermal fields and the Hamiltonian are derived. Due to the time-dependence
of the oscillator frequencies, a counterterm is present in the Hamiltonian. This counterterm determines the
correlation functions in a perturbative fashion. We compute the two point correlation function of the thermal string
at zero order in the power expansion. }

%\begin{center}
%\rule{15cm}{0.01cm}
%\end{center}

\vfill

%\begin{flushright}
%{\footnotesize \textbf{DEFIS-ICE-UFRRJ/2011 \hspace{0.5cm} TH-PHYS/01} }
%\end{flushright}

%-------------------------------------------------------------------------------
\newpage\pagestyle{plain} \pagenumbering{arabic}

\section{Introduction}

Understanding the dynamics of strings in as large a number of spacetime
manifolds as possible represents an essential step in the development of the
string theory and the design of new applications and tests of it. In
particular, one expects to obtain important informations concerning the
semiclassical properties of the gravitational interaction in the string theory
such as the backreaction of the matter to the spacetime background, the
cosmological singularity and evolution and the initial conditions of the
cosmology. Since there are no methods available to explore systematically the
full string physics in arbitrary spacetimes (see \cite{vs} for a review of
some early results), just a few backgrounds in which the above problems can be
addressed consistently within the framework of the available semiclassical
methods have been studied thoroughly.

The manifolds on which it is possible to approach the string dynamics from the
semiclassical point of view have the property of possessing time-like Killing
vectors at least locally. They can be generalized to string backgrounds that
are formed by the time-dependent spacetimes and massless fields from the
closed sector. From these, the distinguished class of time-dependent plane
wave backgrounds have received a considerably attention lately. There are at
least three reasons for which the plane waves are interesting for the string
theory. In the first place, they can be related to cosmological, string and
D-brane backgrounds through the Penrose-G\"{u}ven limit procedure as shown in
\cite{bfhp}. Secondly, the plane waves have the interesting property that they
generate conformal backgrounds in the string theory in which the strings can
be canonically quantized \cite{rt,prt}. And thirdly, the techniques developed
to quantize the strings in plane wave backgrounds can be applied to string
theory in other backgrounds with similar symmetry structure, for example, to
the light-like \cite{cre} and the Kasner-like backgrounds \cite{ckr,kn1,kn2}%
(see for a recent review \cite{ce} and the references therein). The symmetries
of the plane wave backgrounds allows one to study the problems related to the
cosmological singularity in a simpler framework in which the tools of the
perturbative string quantization are available.

In order to fully understand the string dynamics near the cosmological
singularities or the singularities of the time-dependent plane waves, it is
imperative to investigate the thermal aspects of the string theory in the
corresponding time-dependent backgrounds. In the static backgrounds, the
thermal effects can be studied under the hypothesis of the local thermal
equilibrium. Consequently, the thermodynamical quantities of interest can be
derived from the statistical partition function of string. On the other hand,
the assumption of the thermal equilibrium in the time-dependent plane wave
backgrounds leads to the conclusion that the string has inequivalent canonical
representations near the singularity usually taken at zero or infinity time
and at an arbitrary finite value of time, respectively \cite{gmn}. In general,
the necessity of different theories for different values of the time parameter
signals a phase transition in between the two values. However, that is not
displayed by the thermal equilibrium dynamics. Extensive analysis of the
temperature under the thermal equilibrium hypothesis in time-dependent plane
wave backgrounds has been carried out in
\cite{zv,gss,ys1,ys2,blt,gost,hpy,hz,bc} where it has been shown that what is
the equivalent of the Hagedorn temperature in the Minkowski space-time could
be interpreted either as a limiting temperature or a critical temperature or
even an oscillating modes dependent temperature.

The results obtained so far by using different techniques of the quantum
fields at finite temperature point towards the conclusion that the thermal
equilibrium is, at best, a hypothesis valid only locally and asymptotically
away from the singularity. The very general situation in the time-dependent
backgrounds is that the interaction between the gravitational field and the
string involves an exchange of energy at a time dependent rate. As a
consequence, the frequencies of the string modes are time-dependent functions
of time and the string cannot maintain its thermal equilibrium. The
time-dependence of oscillator frequencies lead to sensible effects in the
theory. For example, the correlators of the field theory at finite temperature
have the following general form
%see talk "Nonequilibrium quantum field theory and the lattice" by Jürgen Berges (hep-ph/0409233) page 15.%
\[
G(x,y)=G_{0}(x,y)-\frac{i}{2}\text{sign}_{\mathit{C}}(t_{x}-t_{y})\rho(x,y),
\]
where $G_{0}(x,y)$ is the statistical correlator and $\mathit{C}$ is a closed
real-time contour. At thermal equilibrium, $G_{0}(x,y)\sim\rho(x,y)$ while in
nonequilibrium $G_{0}(x,y)$ is no longer proportional to the spectral function
$\rho(x,y)$. Thus, the heuristic arguments as well as the calculations in the
equilibrium field theory suggest that, in general, the string ensembles in
time-dependent backgrounds are nonequilibrium system.

There are few attempts in the literature to study the nonequilibrium strings
even in the simpler backgrounds, mainly due to the inner difficulties of the
nonequilibrium field theory. However, since the strings are solvable by
canonical methods in the time-dependent plane wave backgrounds one could look
for a formulation of strings in terms of nonequilibrium ensembles of
oscillators. To our knowledge, the few works that address this problem are
doing so in the framework of one approach, known as the Liouville-von Neumann
method \cite{sk,kl} (based on the generalization of the Lewis-Riesenfeld
invariant theorem \cite{lr}) has been used recently to map the time-dependent
quantum mechanics to the Renormalization Group flow and was applied to the
Penrose-G\"{u}ven limit of stacks of D-branes and of the Pilch-Warner solution
in \cite{gzs}. The same method was used to analyze the string spectrum in the
Penrose-G\"{u}ven limit of the NS5-brane plane wave background in \cite{ks}
and to construct the density operator of the type IIB superstring in the
singular scale-invariant plane wave background in \cite{bbl} where it was
shown that the Hagedorn behavior of strings is the same as in the flat
spacetime. However, the results obtained so far are not complete since they
provide a simple description at oscillator level without explaining the
nonequilibrium dynamics of string fields and the way in which the thermal
string observables or correlations should be calculated from it.

In this paper, we propose a new formulation of the nonequilibrium string
dynamics in the time-dependent plane wave backgrounds based on the
Nonequilibrium Thermo Field Dynamics (NETFD) \cite{hu}. The TFD is a real-time
canonical operator formalism in which the statistical averages at
thermodynamical equilibrium are calculated not by performing the explicit
manipulation of the density operator but rather by averaging the corresponding
observables in the thermal vacuum that obeys the thermal state condition that
is defined on the representation space of the TFD (also called thermal space
\cite{hu}). The mathematical structure of the thermal space is that of the
direct product of two identical copies of the Hilbert space of the system at
zero temperature over the field of temperature dependent functions and it is
endowed with an involution denoted by tilde. The physical interpretation of
the thermal vectors is that of thermal degrees of freedom expressed in terms
of tilde and non-tilde quasi-particles. There is a group of linear
involution-invariant and time-invariant maps from the direct product of the
tilde and non-tilde Hilbert spaces at zero temperature to thermal spaces whose
elements are the thermal Bogoliubov operators that preserve the canonical
structure. The thermal group is $SU(1,1)$ for bosons and $SO(2)$ for fermions.
The choice of a certain Bogoliubov transformation is equivalent to the choice
of the thermal vacuum state defined by the corresponding condensate of pairs
of tilde and non-tilde particles. This is so because the Bogoliubov operators
entangle in a specific way the states from the tilde and non-tilde spaces at
zero temperature to produces thermal states \cite{hu}. The equilibrium TFD was
applied to the string theory and the string field theory in flat spacetime in
\cite{yl1,yl2,yl3,yl4,yl5,yl6,fns1,fn1,fn2,fn3,fn4,f1}. More recently, it has
been used to calculate the thermodynamics of strings in static backgrounds in
\cite{ng1,ng2,ng3,ng4,ng5,ivv9,ivv10} and to construct the thermal D-branes
and calculate their thermodynamical functions in
\cite{ivv1,ivv2,ivv3,ivv4,ivv5,mbc1,mbc2} (see for reviews
\cite{ivv6,ivv7,ivv8}). In \cite{ivv11}, the equilibrium TFD was generalized
to accommodate the topological sector of the strings and D-branes on
$R^{1,p}\times T^{d-p-1}$.

The NETFD\ is a canonical real-time formalism in which the nonequilibrium
correlators can be calculated from states and operators that have a simple
algebraic structure. The NETFD formalism retains the basic structure of the
equilibrium TFD with the thermal space being the direct product of two
identical copies of the Hilbert space of the system acted upon by two kinds of
operators with and without tilde, respectively. They form the representation
space of the operator algebra. The thermal vectors are interpreted as thermal
degrees of freedom of the system in nonequilibrium and they can be constructed
in the Fock representation of the thermal space from the thermal vacuum which
is a solution of the master equation%
\begin{equation}
i\frac{\partial}{\partial t}\left\vert 0(t)\right\rangle =\hat{H}\left\vert
0(t)\right\rangle .\label{master-eq}%
\end{equation}
This is the Schroedinger equation with the infinitesimal time evolution
generated by the operator \cite{au1,au2,au3}%
\begin{equation}
\hat{H}=H-\tilde{H},\label{TFD-hamiltonian}%
\end{equation}
where $H$ and $\tilde{H}$ are the Hamiltonians of the non-tilde and tilde
degrees of freedom, respectively. The minus sign is due to the unbounded
energy of the tilde excitations and is a characteristic of the thermal
processes in the TFD formalism \cite{hu}. Also, it shows that in
nonequilibrium the thermal vacuum state is generally unstable and
time-dependent. Nevertheless, one can still define a set of time-dependent
creation and annihilation operators such that the thermal vacuum be
canonically defined for any value of the time parameter \cite{hu}. Since the
time-dependence is carried by the states and the operators, new
time-independent operators are introduced by time-dependent Bogoliubov maps in
order to define stable quasi-particle states. They allow one to calculate the
correlators of the time-dependent thermal operators in the time-independent
quasi-particle vacuum with or without interactions \cite{hu}. The relativistic
generalization of the formalism has been proposed very recently in
\cite{mi,miny}\footnote{While we have been finishing this paper, we learned of
the reference \cite{mi2} in which similar results to ours are obtained for the
complex scalar field in the Minkowski spacetime. The main difference, other
than the context, is the time-depenence of the frequencies of the string
oscillators in contrast to the constant frequencies of the oscillators of the
scalar field from \cite{mi2}.}.

This paper is organized as follows. In Section 2 we briefly review the
canonical structure of the free string in the time-dependent plane wave
background following \cite{prt}. After gauging the world-sheet and the
target-space symmetries, the string can be reduced to an ensemble of
time-dependent non-interacting oscillators in the Brinkmann coordinates. By
exploiting this structure, we construct in Section 3 the time-dependent
Bogoliubov operators and the time-independent quasi-particle vacuum.In Section
4 we construct the time-dependent correlationfunctions of the thermal string.
We show that they contain the transition matrix between the initial time
$t_{0}=-\infty$ and the time of the singularity $t_{1}=0$, respectively. In
NETFD the transition matrix is defined in terms of the interaction Hamiltonian
of the tilde and non-tilde strings and a counterterm that is responsible for
the time evolution of the number operators. A double Dyson-Schwinger expansion
is possible in the two terms. We compute the correlations at zero order of the
expansion. The last section is devoted to conclusions. In order to make the
paper self-consistent, some basic facts about the NETFD formalism are
collected in the Appendix.

\section{Strings in plane wave background}

In this section we review the canonical quantization of the free closed string
in the time-dependent singular plane wave background \cite{prt,ce} and
establish our notations. In what follows, we are going to use the Brinkmann
coordinates in which the background belongs to the following general class
\begin{align}
ds^{2}  &  =2dx^{+}dx^{-}-\lambda\left(  x^{+}\right)  \left(  x^{i}%
dx^{+}\right)  ^{2}+\left(  dx^{i}\right)  ^{2},\label{metric}\\
\phi &  =\phi(x^{+}). \label{dilaton}%
\end{align}
Here, $x^{\pm}$ are the light-cone coordinates in the target-space and
$i,j=2,3,\ldots D$. The dilaton compensates in the Ricci tensor the
contribution of the metric and ensures that the background is conformally
flat. In the case of the\ singular plane waves, the function $\lambda\left(
x\right)  $ behaves as $\lambda\left(  x\right)  \rightarrow kx^{-2}+O\left(
x^{r}\right)  $ at $x\rightarrow0,$ where $k$ is a real parameter and $r>2$.
For $k>0$ the mass is positive and the dilaton takes the form%
\begin{equation}
\phi(x^{+})=\phi_{0}-Cx^{+}+\frac{\left(  D-2\right)  k}{2}\ln(x^{+}),
\label{dilaton-plane-wave}%
\end{equation}
where $C$ is a constant and $D$ is the spacetime dimension. The string action
in curved dilatonic background has the general form \cite{gsw}%
\begin{equation}
S=-\frac{1}{4\pi\alpha^{\prime}}\int d^{2}\sigma\sqrt{-h}\left[
h^{\alpha\beta}\left(  \sigma\right)  g_{\mu\nu}\left(  x\right)
\partial_{\alpha}x^{\mu}\left(  \sigma\right)  \partial_{\beta}x^{\nu}\left(
\sigma\right)  +\frac{\alpha^{\prime}}{2}R^{(2)}\left(  \sigma\right)
\phi\left(  \sigma\right)  \right]  , \label{action}%
\end{equation}
where $\sigma^{\alpha}=(\sigma^{0},\sigma^{1})=(\tau,\sigma)$ are the
world-sheet coordinates, $h^{\alpha\beta}\left(  \sigma\right)  $ is the
world-sheet metric, $x^{\mu}\left(  \sigma\right)  $ are the string fields,
$g_{\mu\nu}\left(  x\right)  $ is the space-time metric defined by the line
element (\ref{metric}) with $\mu,\nu=\pm,2,3,\ldots D$,$\ $and $R^{(2)}\left(
\sigma\right)  $ is the scalar curvature of the world-sheet. The symmetries of
the action (\ref{action}) are fixed by choosing $\partial_{\alpha}%
h_{\alpha\alpha}=0$ and the light-cone gauge $x^{+}=\alpha^{\prime}p^{+}\tau$.

The general solution of the equations of motion obtained from (\ref{action})
has the Fourier expansion \cite{prt}%
\begin{equation}
x^{i}(\tau,\sigma)=x_{0}^{i}(\tau)+i\sqrt{\frac{\alpha^{\prime}}{2}}\sum
_{n>0}\frac{1}{n}\left[  F_{n}(\tau)\left(  \alpha_{n}^{i}e^{2in\sigma}%
+\beta_{n}^{i}e^{-2in\sigma}\right)  -F_{n}^{\ast}(\tau)\left(  \alpha
_{n}^{i\dag}e^{-2in\sigma}+\beta_{n}^{i\dag}e^{2in\sigma}\right)  \right]
.\label{solution-eq-motion}%
\end{equation}
The creation and annihilation operators from $x^{i}(\tau,\sigma)$\ are
normalized as string modes rather than oscillator modes and the following
linear superpositions of the Bessel functions have been introduced%
\begin{align}
F_{n}(\tau) &  =\exp(-\frac{i\pi\nu}{2})\sqrt{n\pi\nu}\left[  J_{\nu-\frac
{1}{2}}\left(  2\pi\tau\right)  -iY_{\nu-\frac{1}{2}}\left(  2\pi\tau\right)
\right]  ,\label{complex-f}\\
G_{n}(\tau) &  =\exp(-\frac{i\pi\nu}{2})\sqrt{n\pi\nu}\left[  J_{\nu+\frac
{1}{2}}\left(  2\pi\tau\right)  -iY_{\nu+\frac{1}{2}}\left(  2\pi\tau\right)
\right]  ,\label{complex-g}%
\end{align}
where $\nu=\left(  1/2\right)  \left(  1+\sqrt{1-4k}\right)  $. The modes of
the string center of mass can be written in terms of time-independent
oscillators as follows%
\begin{equation}
x_{0}^{i}(\tau)=%
%TCIMACRO{\QDATOPD{\{}{.}{\sqrt{\frac{\alpha^{\prime}}{2\left(  2\nu-1\right)
%}}\left[  \left(  \alpha_{0}^{i}+\alpha_{0}^{i\dag}\right)  \tau^{1-\nu
%}-2i\left(  \alpha_{0}^{i}-\alpha_{0}^{i\dag}\right)  \tau^{\nu}\right]
%,k\neq\frac{1}{4},}{\sqrt{\frac{\alpha^{\prime}\tau}{2}}\left[  \left(
%\alpha_{0}^{i}+\alpha_{0}^{i\dag}\right)  -2i\left(  \alpha_{0}^{i}-\alpha
%_{0}^{i\dag}\right)  \ln\tau\right]  ,k=\frac{1}{4}.}}%
%BeginExpansion
\genfrac{\{}{.}{0pt}{0}{\sqrt{\frac{\alpha^{\prime}}{2\left(  2\nu-1\right)
}}\left[  \left(  \alpha_{0}^{i}+\alpha_{0}^{i\dag}\right)  \tau^{1-\nu
}-2i\left(  \alpha_{0}^{i}-\alpha_{0}^{i\dag}\right)  \tau^{\nu}\right]
,k\neq\frac{1}{4},}{\sqrt{\frac{\alpha^{\prime}\tau}{2}}\left[  \left(
\alpha_{0}^{i}+\alpha_{0}^{i\dag}\right)  -2i\left(  \alpha_{0}^{i}-\alpha
_{0}^{i\dag}\right)  \ln\tau\right]  ,k=\frac{1}{4}.}%
%EndExpansion
\label{zero-modes}%
\end{equation}
The Hamiltonian of string in the light-cone gauge has the following oscillator
structure%
\begin{align}
H &  =\frac{\alpha^{\prime}}{2}\sum_{i=2}^{D}\left[  \left(  p_{0}^{i}\right)
^{2}+\frac{k}{4\alpha^{\prime2}\tau^{2}}\left(  x_{0}^{i}\right)  ^{2}\right]
\nonumber\\
&  +\frac{1}{2}\sum_{i=2}^{D}\sum_{n=1}^{\infty}\left[  \Omega_{n}\left(
\tau\right)  \left(  \alpha_{n}^{i\dag}\alpha_{n}^{i}+\beta_{n}^{i\dag}%
\beta_{n}^{i}\right)  -\Phi_{n}\left(  \tau\right)  \alpha_{n}^{i}\beta
_{n}^{i}-\Phi_{n}^{\ast}\left(  \tau\right)  \alpha_{n}^{i\dag}\beta
_{n}^{i\dag}\right]  ,\label{hamiltonian}%
\end{align}
where the time-dependent coefficients $\Omega_{n}\left(  \tau\right)  $ and
$\Phi_{n}(\tau)$ are given by the following relations%
\begin{align}
\Omega_{n}\left(  \tau\right)   &  =\left(  1+\frac{\nu}{4\tau^{2}n^{2}%
}\right)  \left\vert F_{n}(\tau)\right\vert ^{2}+\left\vert G_{n}%
(\tau)\right\vert ^{2}-\frac{\nu}{2n\tau}\left[  F_{n}(\tau)G_{n}^{\ast}%
(\tau)+F_{n}^{\ast}(\tau)G_{n}(\tau)\right]  ,\label{omega}\\
\Phi_{n}(\tau) &  =\left(  1+\frac{\nu}{4\tau^{2}n^{2}}\right)  F_{n}%
(\tau)^{2}+G_{n}(\tau)^{2}-\frac{\nu}{2n\tau}F_{n}(\tau)G_{n}(\tau
).\label{phi}%
\end{align}
From equations (\ref{omega}) and (\ref{phi}), one can see that the string
behaves as a collection of self-interacting time-dependent string modes even
if it is a free string. The non-diagonal interacting terms do not mix modes of
different frequencies. They are generated by the time-dependent couplings
$\Omega_{n}\left(  \tau\right)  $ and $\Phi_{n}(\tau)$ induced by the
time-dependent gravitational field. The self-interacting sector of the
Hamiltonian should be treated non-perturbatively because $\Omega_{n}\left(
\tau\right)  $ and $\Phi_{n}(\tau)$ are of the same strength. However, as was
shown in \cite{prt}, the Hamiltonian can be written as a collection of
non-interacting time-dependent harmonic oscillators by applying the following
linear map%
\begin{align}
A_{n}^{i}\left(  \tau\right)   &  =u_{n}\left(  \tau\right)  \alpha_{n}%
^{i}+v_{n}^{\ast}\left(  \tau\right)  \beta_{n}^{i\dag},\label{map-free-1}\\
A_{n}^{i\dag}\left(  \tau\right)   &  =u_{n}^{\ast}\left(  \tau\right)
\alpha_{n}^{i\dag}+v_{n}\left(  \tau\right)  \beta_{n}^{i},\label{map-free-2}%
\\
B_{n}^{i}\left(  \tau\right)   &  =u_{n}\left(  \tau\right)  \beta_{n}%
^{i}+v_{n}^{\ast}\left(  \tau\right)  \alpha_{n}^{i\dag},\label{map-free-3}\\
B_{n}^{i\dag}\left(  \tau\right)   &  =u_{n}^{\ast}\left(  \tau\right)
\beta_{n}^{i\dag}+v_{n}\left(  \tau\right)  \alpha_{n}^{i},\label{map-free-4}%
\end{align}
where%
\begin{align}
u_{n}\left(  \tau\right)   &  =\frac{1}{2}e^{2i\omega_{n}\left(  \tau\right)
\tau}\left[  F_{n}(\tau)+\frac{i}{2\omega_{n}\left(  \tau\right)  }%
\partial_{\tau}F_{n}(\tau)\right]  ,\label{u-funct}\\
v_{n}\left(  \tau\right)   &  =\frac{1}{2}e^{-2i\omega_{n}\left(  \tau\right)
\tau}\left[  -F_{n}(\tau)+\frac{i}{2\omega_{n}\left(  \tau\right)  }%
\partial_{\tau}F_{n}(\tau)\right]  .\label{v-funct}%
\end{align}
The $A$ and $B$ oscillators satisfy the canonical commutation relations%
\begin{equation}
\left[  A_{n}^{i}\left(  \tau\right)  ,B_{m}^{j\dag}\left(  \tau\right)
\right]  =\left[  A_{n}^{i}\left(  \tau\right)  ,B_{m}^{j\dag}\left(
\tau\right)  \right]  =\delta_{nm}\delta^{ij},\qquad\left[  A_{n}^{i}\left(
\tau\right)  ,B_{m}^{j}\left(  \tau\right)  \right]  =\left[  A_{n}^{i}\left(
\tau\right)  ,B_{m}^{j\dag}\left(  \tau\right)  \right]  =0.\label{free-osc}%
\end{equation}
In terms of the $A$ and $B$ operators, the light-cone Hamiltonian takes the
diagonal form%
\begin{equation}
H=\frac{\alpha^{\prime}}{2}\sum_{i=2}^{D}\left[  \left(  p_{0}^{i}\right)
^{2}+\left(  \frac{k}{4\alpha^{\prime2}\tau^{2}}\right)  ^{2}\left(  x_{0}%
^{i}\right)  ^{2}\right]  +\frac{1}{2}\sum_{i=2}^{D}\sum_{n=1}^{\infty}\left[
\omega_{n}\left(  \tau\right)  \left(  A_{n}^{i\dag}A_{n}^{i}+B_{n}^{i\dag
}B_{n}^{i}\right)  \right]  +h\left(  \tau\right)  ,\label{free-hamiltonian}%
\end{equation}
where the time-dependent frequencies are%
\begin{equation}
\omega_{n}\left(  \tau\right)  =\sqrt{n^{2}+\frac{k}{4\tau^{2}}}%
.\label{free-freq}%
\end{equation}
The time-dependent function%
\begin{equation}
h\left(  \tau\right)  =\left(  D-2\right)  \sum_{n=1}^{\infty}\omega
_{n}\left(  \tau\right)  \label{ord-const}%
\end{equation}
plays the r\^{o}le of the normal-ordering constant. It is logarithmic
divergent and it can be cancelled by a field renormalization of the dilaton
for all values of $\tau$ \cite{prt}. The string fields
(\ref{solution-eq-motion})\ can be expanded in terms of $A$ and $B$
oscillators as follows%
\begin{align}
x^{i}(\tau,\sigma) &  =x_{0}^{i}(\tau)+i\sqrt{\frac{\alpha^{\prime}}{2}}%
\sum_{n>0}\frac{1}{\sqrt{n\omega_{n}\left(  \tau\right)  }}\left\{
e^{-2i\omega_{n}\left(  \tau\right)  \tau}A_{n}^{i}\left(  \tau\right)
e^{2in\sigma}-e^{2i\omega_{n}\left(  \tau\right)  \tau}A_{n}^{i\dag}\left(
\tau\right)  e^{-2in\sigma}\right\}  \label{free-sol-eq-mot}\\
&  +i\sqrt{\frac{\alpha^{\prime}}{2}}\sum_{n>0}\frac{1}{\sqrt{n\omega
_{n}\left(  \tau\right)  }}\left\{  e^{-2i\omega_{n}\left(  \tau\right)  \tau
}B_{n}^{i}\left(  \tau\right)  e^{-2in\sigma}-e^{2i\omega_{n}\left(
\tau\right)  \tau}B_{n}^{i\dag}\left(  \tau\right)  e^{2in\sigma}\right\}
.\nonumber
\end{align}
In general, the particle-like states of time-dependent oscillators are well
defined only asymptotically \cite{hu}. In the present case, there are two
asymptotic Fock spaces of string oscillators for observers at $\tau
\rightarrow\pm\infty$. Since the oscillator sector is asymptotically the same
as that of the free string in the flat space-time, the asymptotic states can
be constructed from the light-cone vacua $\left\vert 0(\pm\infty)\right\rangle
_{\alpha-\beta}$ which are eigenstates of $\alpha_{n}^{i}$ and $\beta_{n}^{i}$
operators. At finite $\tau$, the free string picture is lost due to the
self-interactions. However, there are free states that can be obtained by
exciting the time-dependent vacuum $\left\vert 0(\tau)\right\rangle _{A-B}$ in
the representation defined by the operators $A$ and $B$, respectively. The
dynamics of the center-of-mass of string does not reduce to the corresponding
one in flat space-time, and thus is characteristic to the time-dependent plane
wave background.

\section{Nonequilibrium dynamics of strings}

The fact that the string theory in the time-dependent plane wave backgrounds
is solvable in the canonical quantization suggests that the nonequilibrium
dynamics could have a canonical formulation, too. There are several methods
that can be used to this end. As mentioned in the introduction, we are going
to formulate the nonequilibrium dynamics in the NETFD framework because of the
simple interpretation that can be given to the states, the operators and the
nonequilibrium correlators.

\subsection{Thermal string fields}

The first step to be taken in order to develop the NETFD method for strings is
to define the thermal degrees of freedom. The thermal states belong to the
thermal space $\widehat{\mathcal{H}}=\mathcal{H}\otimes\mathcal{\tilde{H}}$
which is the direct product of two identical copies of the Hilbert space of
string.\ On $\widehat{\mathcal{H}}$ act operators of the form $O\otimes
\tilde{1}$ and $1\otimes\tilde{O}$ that can be conveniently tensored with
vectors from $%
%TCIMACRO{\U{211d} }%
%BeginExpansion
\mathbb{R}
%EndExpansion
^{2}$ to form the so called \emph{thermal doublets} \cite{hu}. We start with
the diagonal form of the theory in which the string fields are given by the
relation (\ref{free-sol-eq-mot}). The coordinates of the tilde string are
obtained by applying the tilde operation to $x^{i}(\tau,\sigma)$ (see the
Appendix)%
\begin{align}
\tilde{x}^{i}(\tau,\sigma) &  =\tilde{x}_{0}^{i}(\tau)-i\sqrt{\frac
{\alpha^{\prime}}{2}}\sum_{n>0}\frac{1}{\sqrt{n\omega_{n}\left(  \tau\right)
}}\left\{  e^{2i\omega_{n}\left(  \tau\right)  \tau}\tilde{A}_{n}^{i}\left(
\tau\right)  e^{-2in\sigma}-e^{-2i\omega_{n}\left(  \tau\right)  \tau}%
\tilde{A}_{n}^{i\dag}\left(  \tau\right)  e^{2in\sigma}\right\}  \nonumber\\
&  -i\sqrt{\frac{\alpha^{\prime}}{2}}\sum_{n>0}\frac{1}{\sqrt{n\omega
_{n}\left(  \tau\right)  }}\left\{  e^{2i\omega_{n}\left(  \tau\right)  \tau
}\tilde{B}_{n}^{i}\left(  \tau\right)  e^{2in\sigma}-e^{-2i\omega_{n}\left(
\tau\right)  \tau}\tilde{B}_{n}^{i\dag}\left(  \tau\right)  e^{-2in\sigma
}\right\}  .\label{free-sol-eq-mot-til}%
\end{align}
It is convenient to absorb the time-dependent exponentials into the canonical
operators by the mean of the following change of variables%
\begin{equation}
e^{-2i\omega_{n}\left(  \tau\right)  \tau}A_{n}^{i}\left(  \tau\right)
=a_{n}^{i}\left(  \tau\right)  ,\qquad e^{2i\omega_{n}\left(  \tau\right)
\tau}A_{n}^{i\dag}\left(  \tau\right)  =a_{n}^{i\dag}\left(  \tau\right)
,\label{a-op}%
\end{equation}
with similar redefinition of operators in other sectors. The string thermal
doublet, denoted by $\phi^{i\alpha}(\tau,\sigma)$, can be obtained by
organizing the fields from (\ref{free-sol-eq-mot}) and
(\ref{free-sol-eq-mot-til}) in to a two dimensional vector field in such a way
that the $a-b$ structure be maintained%
\begin{align}
\phi^{i\alpha}(\tau,\sigma) &  =\phi_{0}^{i\alpha}(\tau)+\phi_{a}^{i\alpha
}(\tau,\sigma)+\phi_{b}^{i\alpha}(\tau,\sigma)\nonumber\\
&  =\phi_{0}^{i\alpha}(\tau)+i\sqrt{\frac{\alpha^{\prime}}{2}}\sum_{n>0}%
\frac{1}{\sqrt{n\omega_{n}\left(  \tau\right)  }}\left\{  a_{n}^{i\alpha
}\left(  \tau\right)  e^{2in\sigma}-\left(  s_{3}\bar{a}_{n}^{i}\left(
\tau\right)  ^{T}\right)  ^{\alpha}e^{-2in\sigma}\right\}  \nonumber\\
&  -i\sqrt{\frac{\alpha^{\prime}}{2}}\sum_{n>0}\frac{1}{\sqrt{n\omega
_{n}\left(  \tau\right)  }}\left\{  b_{n}^{i\alpha}\left(  \tau\right)
e^{-2in\sigma}-\left(  s_{3}\bar{b}_{n}^{i}\left(  \tau\right)  ^{T}\right)
^{\alpha}e^{2in\sigma}\right\}  ,\label{free-sol-eq-mot-bar}%
\end{align}
where $\alpha$ is the $%
%TCIMACRO{\U{211d} }%
%BeginExpansion
\mathbb{R}
%EndExpansion
^{2}$ index, $\phi_{0}^{i\alpha}(\tau)$ stands for the thermal doublet of the
center of mass and $s_{3}$ is the Pauli matrix. Also, we have introduced the
standard $%
%TCIMACRO{\U{211d} }%
%BeginExpansion
\mathbb{R}
%EndExpansion
^{2}$ thermal doublet%
\begin{equation}
a_{n}^{i\alpha}\left(  \tau\right)  =%
\begin{pmatrix}
a_{n}^{i}\left(  \tau\right)  \\
\tilde{a}_{n}^{i\dag}\left(  \tau\right)
\end{pmatrix}
,\qquad\bar{a}_{n}^{i\alpha}\left(  \tau\right)  =%
\begin{pmatrix}
a_{n}^{i\dag}\left(  \tau\right)   & -\tilde{a}_{n}^{i}\left(  \tau\right)
\end{pmatrix}
.\label{thermal-a-op}%
\end{equation}
Similar operators can be introduced in the $b$ - sector. The conjugate of the
thermal field given by the relation (\ref{free-sol-eq-mot-bar}) is obtained by
applying the operations defined in the Appendix%
\begin{align}
\bar{\phi}^{i\alpha}(\tau,\sigma) &  =\bar{\phi}_{0}^{i\alpha}(\tau)+\bar
{\phi}_{a}^{i\alpha}(\tau,\sigma)+\bar{\phi}_{b}^{i\alpha}(\tau,\sigma
)\nonumber\\
&  =\bar{\phi}_{0}^{i\alpha}(\tau)-i\sqrt{\frac{\alpha^{\prime}}{2}}\sum
_{n>0}\frac{1}{\sqrt{n\omega_{n}\left(  \tau\right)  }}\left\{  \bar{a}%
_{n}^{i\alpha}\left(  \tau\right)  e^{-2in\sigma}-\left(  a_{n}^{i}\left(
\tau\right)  ^{T}s_{3}\right)  ^{\alpha}e^{2in\sigma}\right\}  \nonumber\\
&  +i\sqrt{\frac{\alpha^{\prime}}{2}}\sum_{n>0}\frac{1}{\sqrt{n\omega
_{n}\left(  \tau\right)  }}\left\{  \bar{b}_{n}^{i\alpha}\left(  \tau\right)
e^{2in\sigma}-\left(  b_{n}^{i}\left(  \tau\right)  ^{T}s_{3}\right)
^{\alpha}e^{-2in\sigma}\right\}  .\label{free-sol-eq-mot-bar-bar}%
\end{align}
Note that the coordinates of the center of mass do not have the standard form
of a thermal doublet as given by the NETFD. This can be remedied by organizing
the coordinates as follows%
\begin{equation}
x_{0}^{i}(\tau)=U(\tau)\alpha_{0}^{i}+U^{\ast}(\tau)\alpha_{0}^{i\dag
},\label{coord-cm-1}%
\end{equation}
where the time-dependent coefficients are given by the following relations%
\begin{equation}
U(\tau)=\left\{
\begin{array}
[c]{cc}%
\sqrt{\frac{\alpha^{\prime}}{2\left(  2\nu-1\right)  }}\tau^{1-\nu}%
-i\sqrt{\frac{2\alpha^{\prime}}{2\nu-1}}\tau^{\nu}, & k\neq\frac{1}{4}\\
\sqrt{\frac{\alpha^{\prime}\tau}{2}}-i\sqrt{2\alpha^{\prime}\tau}\ln\tau, &
k=\frac{1}{4}%
\end{array}
\right.  .\label{U-funct}%
\end{equation}
The coordinates of the center of mass of the tilde string can be obtained by
applying the conjugation rules given in the Appendix. Then the thermal fields
are generalized to the center of mass by the following relations%
\begin{equation}
\phi_{0}^{i\beta}(\tau)=%
\begin{pmatrix}
x_{0}^{i}(\tau)\\
\tilde{x}_{0}^{i}(\tau)
\end{pmatrix}
=U(\tau)\alpha_{0}^{i\beta}+U^{\ast}(\tau)\left(  s_{3}\bar{\alpha}_{0}%
^{iT}\right)  ^{\beta}.\label{coord-cm-2}%
\end{equation}
The thermal doublets define a time-parametrized family of representations of
the thermal field constructed from the thermal vacua $\left\{  \left\vert
0(\tau)\right\rangle \right\}  $. Any of these representations can be mapped
to the time-independent quasi-particle representation by the inverse of the
Bogoliubov map (see the Appendix)%
\begin{equation}
a_{n}^{i\alpha}\left(  \tau\right)  =\mathbf{B}_{a,n}^{-1}(\tau)^{\alpha\beta
}\xi_{n}^{\beta},\qquad b_{n}^{i\alpha}\left(  \tau\right)  =\mathbf{B}%
_{b,n}^{-1}(\tau)^{\alpha\beta}\chi_{n}^{\beta}.\label{Bog-map}%
\end{equation}
Then the nonequilibrium dynamics is determined by the time variation of the
thermal doublets $a_{n}^{i\alpha}\left(  \tau\right)  $ and $b_{n}^{i\alpha
}\left(  \tau\right)  $ in the time-independent vacuum $\left\vert
0\right\rangle $ defined by $\xi_{n}^{\alpha}$ and $\chi_{n}^{\alpha}$,
respectively. The Bogoliubov operators can be different in the $a$ and $b$
sectors, respectively, but they are the same for all transversal directions.
However, since the two sectors differ only in the orientation of the modes
along the space-like direction of the world-sheet, we can take $\mathbf{B}%
_{a,n}(\tau)=\mathbf{B}_{b,n}(\tau)=\mathbf{B}_{n}(\tau)\footnote{In the case
of the complex scalar field discussed in \cite{mi2} this equality does not
hold because the left and right moving states have different charges.}$. It
follows that the Bogoliubov map of the $k$ - mode has the form given by the
relation (\ref{A-16}) in the linear thermal gauge%
\begin{align}
\xi_{k}^{j\alpha} &  =\exp\left[  i\int_{\tau_{0}}^{\tau}d\lambda\omega
_{k}(\lambda)\right]  B_{k}^{\alpha\beta}\left(  \tau\right)  a_{k}^{j\beta
}\left(  \tau\right)  ,\label{Bog-op-a}\\
\bar{\xi}_{k}^{j\alpha} &  =\bar{a}_{k}^{j\beta}\left(  \tau\right)
\exp\left[  -i\int_{\tau_{0}}^{\tau}d\lambda\omega_{k}(\lambda)\right]
\left[  B_{k}^{-1}\right]  ^{\beta\alpha}\left(  \tau\right)
,\label{Bog-op-b}\\
\chi_{k}^{j\alpha} &  =\exp\left[  i\int_{\tau_{0}}^{\tau}d\lambda\omega
_{k}(\lambda)\right]  B_{k}^{\alpha\beta}\left(  \tau\right)  b_{k}^{j\beta
}\left(  \tau\right)  ,\label{Bog-op-c}\\
\bar{\chi}_{k}^{j\alpha} &  =\bar{b}_{k}^{j\beta}\left(  \tau\right)
\exp\left[  -i\int_{\tau_{0}}^{\tau}d\lambda\omega_{k}(\lambda)\right]
\left[  B_{k}^{-1}\right]  ^{\beta\alpha}\left(  \tau\right)
,\label{Bog-op-d}%
\end{align}
where the positive exponential represents the pure complex phase function and%
\begin{equation}
n_{k}(\tau)\delta_{k,l}=n_{k}^{i}(\tau)\delta_{k,l}=\left\langle
0(\tau)\left\vert a_{k}^{i\dag}a_{l}^{i}\right\vert 0(\tau)\right\rangle
.\label{n-k}%
\end{equation}
The initial boundary conditions are taken at $\tau_0\rightarrow-\infty$. The
time-dependent quasi-particle operators are obtained by multiplying the
transformations given in the equations (\ref{Bog-op-a}) - (\ref{Bog-op-d})
with the inverse of the corresponding phase functions to obtain%
\begin{align}
\xi_{k}^{i\alpha}\left(  \tau\right)   &  =B_{k}^{\alpha\beta}\left(
\tau\right)  a_{k}^{i\beta}\left(  \tau\right)  ,\label{quasi-part-1}\\
\bar{\xi}_{k}^{i\alpha}\left(  \tau\right)   &  =\bar{a}_{k}^{i\beta}\left(
\tau\right)  \left[  B_{k}^{-1}\right]  ^{\beta\alpha}\left(  \tau\right)
,\label{quasi-part-2}\\
\chi_{k}^{i\alpha}\left(  \tau\right)   &  =B_{k}^{\alpha\beta}\left(
\tau\right)  b_{k}^{i\beta}\left(  \tau\right)  ,\label{quasi-part-3}\\
\bar{\chi}_{k}^{i\alpha}\left(  \tau\right)   &  =\bar{b}_{k}^{i\beta}\left(
\tau\right)  \left[  B_{k}^{-1}\right]  ^{\beta\alpha}\left(  \tau\right)
.\label{quasi-part-4}%
\end{align}
The quasi-particle representation is based on the oscillator equation of
motion that is satisfied by the string modes (\ref{solution-eq-motion}) and by
the natural assumption that the system of oscillators evolves according to the
Schroedinger equation with the Hamiltonian (\ref{free-hamiltonian}) in the
$a-b$ representation \cite{prt}. By taking the first derivative of $%
\xi_k^i\alpha\left(  \tau\right)  $ with respect with $\tau$ one can easily
check that $a_k^j\beta\left(  \tau\right)  $ satisfies the first order
differential equation%
\begin{equation}
\left[  \left(  i\frac{d}{d\tau}-\omega_{k}(\tau)\right)  +P_{k}(\tau)\right]
^{\alpha\beta}a_{k}^{j\beta}\left(  \tau\right)  =0,\label{evol-a}%
\end{equation}
where%
\begin{equation}
P_{k}^{\alpha\beta}(\tau)=i\frac{dn_{k}(\tau)}{d\tau}%
\begin{pmatrix}
1 & -1\\
1 & 1
\end{pmatrix}
.\label{P-op}%
\end{equation}
The same equations hold for $b_k^j\beta\left(  \tau\right)  $ operators. This
is a general result of NETFD for the set of time-dependent oscillators
\cite{hu}. From (\ref{evol-a}) it follows that the time-evolution of the
thermal string modes obeys the following equations%
\begin{align}
i\frac{d}{d\tau}a_{k}^{j\alpha}\left(  \tau\right)   &  =\left[
a_{k}^{j\alpha}\left(  \tau\right)  ,\hat{H}_{Qa}\right]
,\label{eq-mot-therm-1}\\
i\frac{d}{d\tau}\bar{a}_{k}^{j\alpha}\left(  \tau\right)   &  =\left[  \bar
{a}_{k}^{j\alpha}\left(  \tau\right)  ,\hat{H}_{Qa}\right]
,\label{eq-mot-therm-2}%
\end{align}
where%
\begin{equation}
\hat{H}_{Q}^{a}(\tau)=\sum_{i=2}^{D}\sum_{k}\left[  \omega_{k}(\tau
)\delta^{^{\alpha\beta}}+P_{k}^{\alpha\beta}(\tau)\right]  \bar{a}%
_{k}^{i\alpha}\left(  \tau\right)  a_{k}^{i\beta}\left(  \tau\right)
.\label{therm-ham}%
\end{equation}
The second term in the equation (\ref{therm-ham}) is the thermal counterterm
which should be added to the Hamiltonian as a consequence of the
time-dependent Bogoliubov map. As can be seen from its definition (\ref{P-op})
it is related to the variation of $n_k(\tau)$ from the relation (\ref{n-k}).
The same relations holds in the $b$ - sector. Since the total Hamiltonian can
be written as \cite{hu}%
\begin{equation}
\hat{H}=\hat{H}_{0}+\hat{H}_{int}=\hat{H}_{Q}+\hat{H}_{I},\label{H-total-int}%
\end{equation}
the counterterm $\hat{H}_Q(\tau)=\hat{H}_cm(\tau)+\hat{H}_Q^a(\tau)+\hat
{H}_Q^b(\tau)$ from (\ref{H-total-int}) should be compensated by a term to be
added to the interaction term. In the case of the free theory, the interaction
Hamiltonian contains only the compensator of the counterterm. Note that the
quasi-particle fields, even if stable, cannot be used to define the asymptotic
thermal states because the thermal Hamiltonian is unbounded from below
\cite{hu}.

\subsection{Dynamics of thermal fields}

Once the time-evolution of each mode of the thermal string, described by the
equations (\ref{eq-mot-therm-1}) and (\ref{eq-mot-therm-2}), is understood, we
can proceed to deriving the equations of motion of the thermal fields. Since
we are working in the Hamiltonian formalism, we need to determine the
canonical conjugates of $\phi^{i\alpha}(\tau,\sigma)$\ and $\bar{\phi
}^{i\alpha}(\tau,\sigma)$. The conjugate momenta are defined such that they
satisfy the equal-time commutation relations \cite{hu}%
\begin{align}
\left[  \phi^{i\alpha}(\tau,\sigma),\pi^{j\beta}(\tau,\sigma^{\prime})\right]
&  =i\delta^{ij}\delta(\sigma-\sigma^{\prime})s_{3}^{\alpha\beta
},\label{eq-t-comm-1}\\
\left[  \bar{\phi}^{i\alpha}(\tau,\sigma),\pi^{j\beta}(\tau,\sigma^{\prime
})\right]   &  =i\delta^{ij}\delta(\sigma-\sigma^{\prime})\delta^{\alpha\beta
}.\label{eq-t-comm-2}%
\end{align}
By using the (\ref{eq-mot-therm-1}) and (\ref{eq-mot-therm-2}) in the above
commutators we obtain%
\begin{align}
\pi^{i\alpha}(\tau,\sigma) &  =\pi_{0}^{i\alpha}(\tau)+\pi_{a}^{i\alpha}%
(\tau,\sigma)+\pi_{b}^{i\alpha}(\tau,\sigma)\nonumber\\
&  =\frac{i}{2\pi}\left[  U^{-1}(\tau)\left(  s_{3}\bar{\alpha}_{0}%
^{i}\right)  ^{\alpha}+\left(  U^{\ast}(\tau)\right)  ^{-1}\alpha_{0}%
^{i\alpha}\right]  \nonumber\\
&  +\frac{i}{\sqrt{2\alpha^{\prime}}}\sum_{n>0}\sqrt{n\omega_{n}\left(
\tau\right)  }\left\{  a_{n}^{i\alpha}\left(  \tau\right)  e^{2in\sigma
}+\left(  s_{3}\bar{a}_{n}^{i}\left(  \tau\right)  ^{T}\right)  ^{\alpha
}e^{-2in\sigma}\right\}  \nonumber\\
&  -\frac{i}{\sqrt{2\alpha^{\prime}}}\sum_{n>0}\sqrt{n\omega_{n}\left(
\tau\right)  }\left\{  b_{n}^{i\alpha}\left(  \tau\right)  e^{-2in\sigma
}+\left(  s_{3}\bar{b}_{n}^{i}\left(  \tau\right)  ^{T}\right)  ^{\alpha
}e^{2in\sigma}\right\}  .\label{pi-mom}%
\end{align}
The conjugate momenta of the fields $\bar{\phi}^{i\alpha}(\tau,\sigma)$ can be
obtained by applying the tilde conjugation to $\pi^{i\alpha}(\tau,\sigma)$ and
they take the following form%
\begin{align}
\bar{\pi}^{i\alpha}(\tau,\sigma) &  =\bar{\pi}_{0}^{i\alpha}(\tau)+\bar{\pi
}_{a}^{i\alpha}(\tau,\sigma)+\bar{\pi}_{b}^{i\alpha}(\tau,\sigma)\nonumber\\
&  =\frac{i}{2\pi}\left[  -\left(  U^{\ast}(\tau)\right)  ^{-1}\left(
\bar{\alpha}_{0}^{iT}s_{3}\right)  ^{\alpha}+U^{-1}(\tau)\bar{\alpha}%
_{0}^{i\alpha}\right]  \nonumber\\
&  +\frac{i}{\sqrt{2\alpha^{\prime}}}\sum_{n>0}\sqrt{n\omega_{n}\left(
\tau\right)  }\left\{  \bar{a}_{n}^{i\alpha}\left(  \tau\right)  e^{2in\sigma
}+\left(  \bar{a}_{n}^{i}\left(  \tau\right)  ^{T}s_{3}\right)  ^{\alpha
}e^{-2in\sigma}\right\}  \nonumber\\
&  -\frac{i}{\sqrt{2\alpha^{\prime}}}\sum_{n>0}\sqrt{n\omega_{n}\left(
\tau\right)  }\left\{  \bar{b}_{n}^{i\alpha}\left(  \tau\right)
e^{-2in\sigma}+\left(  \bar{b}_{n}^{i}\left(  \tau\right)  ^{T}s_{3}\right)
^{\alpha}e^{2in\sigma}\right\}  .\label{pi-bar-mom}%
\end{align}
The fields obtained in this way satisfy the commutation relations%
\begin{equation}
\left[  \bar{\phi}^{i\alpha}(\tau,\sigma),\bar{\pi}^{j\beta}(\tau
,\sigma^{\prime})\right]  =i\delta^{ij}\delta(\sigma-\sigma^{\prime}%
)\delta^{\alpha\beta},\label{eq-t-comm-3}%
\end{equation}
which qualifies them for the canonical momenta of the $\bar{\phi}^{i\alpha
}(\tau,\sigma)$ fields.

Let us determine the equations of motion of the oscillator sector of the
fields $\phi^{i\alpha}(\tau,\sigma),\bar{\phi}^{i\alpha}(\tau,\sigma
),\pi^{i\alpha}(\tau,\sigma)$ and $\bar{\pi}^{i\alpha}(\tau,\sigma)$. They can
be obtained by deriving the fields with respect to $\tau$ and by using the
equations (\ref{eq-mot-therm-1}) and (\ref{eq-mot-therm-2}). Before doing the
calculation, it is convenient to express the oscillator frequencies in string
units through the rescalling%
\begin{equation}
\omega_{n}\left(  \tau\right)  \rightarrow\frac{n\omega_{n}\left(
\tau\right)  }{\alpha^{\prime}}.\label{rescale-freq}%
\end{equation}
Lengthy calculations give the following equations of motion for the
oscillating modes%
\begin{align}
\left[  \left(  1+2i\partial_{\tau}n_{|\nabla|}\nabla^{-1}T_{0}\right)
^{\alpha\beta}\partial_{\tau}+\partial_{\tau}\nabla\cdot\nabla^{-1}\left(
1+2i\partial_{\tau}n_{|\nabla|}\nabla^{-1}T_{0}\right)  ^{\alpha\beta}\right]
\phi_{osc}^{i\beta}(\tau,\sigma) &  =\pi_{osc}^{i\alpha}(\tau,\sigma
),\label{eq-mot-phi-soc}\\
\left(  1-2i\partial_{\tau}n_{|\nabla|}\nabla^{-1}T_{0}\right)  ^{\alpha\beta
}\nabla^{2}\phi_{osc}^{i\beta}(\tau,\sigma)+\left[  \partial_{\tau}%
-\partial_{\tau}\nabla\cdot\nabla^{-1}\right]  \pi_{osc}^{i\alpha}(\tau
,\sigma) &  =0.\label{eq-mot-pi-osc}%
\end{align}
Here, we have introduced the notation $\nabla=\sqrt{-\partial_{\sigma}%
^{2}+\frac{k}{\tau^{2}}}$. The equations (\ref{eq-mot-phi-soc}) and
(\ref{eq-mot-pi-osc}) describe the classical dynamics of the oscillator sector
of string fields near the plane wave singularity. Since the equations of
motion are the result of the time evolution generated by the classical
Hamiltonian, one can use (\ref{eq-mot-phi-soc}), (\ref{eq-mot-pi-osc}) and
(\ref{eq-mot-cm}) to integrate the Hamilton equations. The resulting
Hamiltonian is composed from two terms%
\begin{equation}
\hat{H}_{Q}=\hat{H}_{cm}+\hat{H}_{osc},\label{hamilt-thermal}%
\end{equation}
where $\hat{H}_{cm}$ is the Hamiltonian of the center of mass and%
\begin{align}
\hat{H}_{osc} &  =\frac{1}{2}\int d\sigma\sum_{i=2}^{D}\sum_{n>0}\left[
\bar{\pi}_{osc}^{i\alpha}\left(  1+2i\partial_{\tau}n_{|\nabla|}\nabla
^{-1}T_{0}\right)  ^{\alpha\beta}\pi_{osc}^{i\beta}\right.  \nonumber\\
&  +\bar{\phi}^{i\alpha}(\tau,\sigma)\left(  1-2i\partial_{\tau}n_{|\nabla
|}\nabla^{-1}T_{0}\right)  ^{\alpha\beta}\nabla^{2}\phi_{osc}^{i\beta}%
(\tau,\sigma)\nonumber\\
&  -\bar{\pi}_{osc}^{i\alpha}\left(  1+T_{n}\right)  ^{\alpha\beta}%
\partial_{\tau}\nabla\cdot\nabla^{-1}\left(  1+2i\partial_{\tau}n_{|\nabla
|}\nabla^{-1}T_{0}\right)  ^{\beta\gamma}\phi_{osc}^{i\gamma}(\tau
,\sigma)\nonumber\\
&  -\left.  \bar{\phi}^{i\alpha}(\tau,\sigma)\partial_{\tau}\nabla\cdot
\nabla^{-1}\pi_{osc}^{i\alpha}\right]  ,\label{H-osc-therm}%
\end{align}
where
\begin{equation}
T_{0}=%
\begin{pmatrix}
1 & -1\\
1 & -1
\end{pmatrix}
,\quad T_{n}=%
\begin{pmatrix}
2i\partial_{\tau}n_{|\nabla|}\nabla^{-1} & 1-2i\partial_{\tau}n_{|\nabla
|}\nabla^{-1}\\
2i\partial_{\tau}n_{|\nabla|}\nabla^{-1} & 1-2i\partial_{\tau}n_{|\nabla
|}\nabla^{-1}%
\end{pmatrix}
.\label{T-matrices}%
\end{equation}
The oscillator Hamiltonian can be split in two terms: the free oscillator
$\hat{H}_{osc}^{0}$\ term and the counterterm $\hat{Q}$ which can be both read
off of the right hand side of (\ref{H-osc-therm}). The Hamiltonian
(\ref{hamilt-thermal}) generates the equations of motion (\ref{eq-mot-phi-soc}%
), (\ref{eq-mot-pi-osc}) and (\ref{eq-mot-cm}). The counterterm $\hat{Q}$
provides the dynamics of the number function $n_{|\nabla|}(\tau)$ from which
is derived the time-evolution of each oscillator mode.

\subsection{Dynamics of center of mass}

Note that the derivation of the oscillator terms from (\ref{pi-mom}) and
(\ref{pi-bar-mom}) follows from the standard NETFD method. However, the
coefficient of the zero mode component $\pi_{0}^{i\alpha}(\tau)$ is not given
by the general formalism. In order to obtain the correct factor one has to
generalize the NETFD in the following way. Firstly, we consider the series
representation of the delta-function%
\begin{equation}
\delta(\sigma-a)=\frac{1}{2\pi}%
%TCIMACRO{\dsum \limits_{n=-\infty}^{\infty}}%
%BeginExpansion
{\displaystyle\sum\limits_{n=-\infty}^{\infty}}
%EndExpansion
e^{2in(\sigma-a)}.\label{delta-series}%
\end{equation}
Then one can show that the commutation relations (\ref{eq-t-comm-1}) and
(\ref{eq-t-comm-2}) are reproduced if the coordinates of the center of mass
satisfy the following commutators%
\begin{equation}
\left[  \phi_{0}^{i\alpha}(\tau),\pi_{0}^{j\beta}(\tau)\right]  =-\frac{i}%
{\pi}\delta^{ij}s_{3}^{\alpha\beta}.\label{cm-comm}%
\end{equation}
From that we can determine the momentum which has the form%
\begin{equation}
\pi_{0}^{i\alpha}(\tau)=\frac{i}{2\pi}\left[  U^{-1}(\tau)\left(  s_{3}%
\bar{\alpha}_{0}^{i}\right)  ^{\alpha}+\left(  U^{\ast}(\tau)\right)
^{-1}\alpha_{0}^{i\alpha}\right]  .\label{zero-mom}%
\end{equation}
This fixes the constant in front of the zero mode term from equation
(\ref{pi-mom}). The coefficient of the corresponding term from the equation
(\ref{pi-bar-mom}) can be obtained in the same way.

The equations of motion of the center of mass can be obtained by taking the
time derivative of the equation (\ref{coord-cm-2}). After some algebra one
obtains the following equation%
\begin{equation}
\left[  \partial_{\tau}+\frac{2\pi}{U^{\ast}(\tau)}\left(  \frac
{\partial_{\tau}U(\tau)}{2\pi-i}-\frac{i\partial_{\tau}U^{\ast}(\tau)}%
{2\pi(2\pi+i)}\right)  \right]  \phi_{0}^{i\alpha}(\tau)=2\pi U(\tau)\left(
\frac{\partial_{\tau}U(\tau)}{2\pi-i}-\frac{i\partial_{\tau}U^{\ast}(\tau
)}{2\pi+i}\right)  \pi_{0}^{i\alpha}.\label{eq-mot-cm}%
\end{equation}
The equation of motion of the momenta is%
\begin{align}
\left[  \partial_{\tau}-i\left(  \frac{1}{2\pi+i}\frac{\partial_{\tau}U(\tau
)}{U(\tau)}+\frac{1}{2\pi-i}\frac{i\partial_{\tau}U^{\ast}(\tau)U(\tau
)}{\left[  U^{\ast}(\tau)\right]  ^{2}}\right)  \right]    & \pi_{0}^{i\alpha
}=\nonumber\\
& -i\left[  \frac{i}{2\pi\left(  2\pi+i\right)  }\frac{\partial_{\tau}U(\tau
)}{U(\tau)\left\vert U(\tau)\right\vert ^{2}}-\frac{1}{2\pi-i}\frac
{\partial_{\tau}U^{\ast}(\tau)}{\left(  U^{\ast}(\tau)\right)  ^{3}}\right]
\phi_{0}^{i\alpha}(\tau).\label{eq-mot-cm-1}%
\end{align}
The equations of motion of $\bar{\phi}_{0}^{i\alpha}(\tau)$ and $\bar{\pi}%
_{0}^{i\alpha}(\tau)$ can be obtained in the same way. From the requirement
that the equations of motion be the Hamilton equations for the Hamiltonian of
the center of mass $\hat{H}_{cm}$, one can calculate the explicit form of it%
\begin{align}
\hat{H}_{cm} &  =\sum_{i=2}^{D}\left[  2\pi U(\tau)\left(  \frac
{\partial_{\tau}U(\tau)}{2\pi-i}+\frac{\partial_{\tau}U^{\ast}(\tau)}{2\pi
+i}\right)  \bar{\pi}_{0}^{i\alpha}\pi_{0}^{i\alpha}+\right.  \nonumber\\
&  -i\left(  \frac{i}{2\pi\left(  2\pi+i\right)  }\frac{\partial_{\tau}%
U(\tau)}{U(\tau)\left\vert U(\tau)\right\vert ^{2}}-\frac{1}{2\pi-i}%
\frac{\partial_{\tau}U^{\ast}(\tau)}{\left(  U^{\ast}(\tau)\right)  ^{3}%
}\right)  \bar{\phi}_{0}^{i\alpha}(\tau)\phi_{0}^{i\alpha}(\tau)\nonumber\\
&  -i\left(  \frac{1}{2\pi+i}\frac{\partial_{\tau}U(\tau)}{U(\tau)}+\frac
{1}{2\pi-i}\frac{i\partial_{\tau}U^{\ast}(\tau)U(\tau)}{\left[  U^{\ast}%
(\tau)\right]  ^{2}}\right)  \bar{\phi}_{0}^{i\alpha}(\tau)\pi_{0}^{i\alpha
}\nonumber\\
&  \left.  -i\frac{2\pi}{U^{\ast}(\tau)}\left(  \frac{\partial_{\tau}U(\tau
)}{2\pi-i}-\frac{i\partial_{\tau}U^{\ast}(\tau)}{2\pi(2\pi+i)}\right)
\bar{\pi}_{0}^{i\alpha}\phi_{0}^{i\alpha}\right]  .\label{H-cm-thermal}%
\end{align}

The total Hamiltonian $\hat{H}_{cm}+\hat{H}_{osc}$ determine the dynamics of
the thermal string field in the proximity of the singularity.

\section{Thermal string correlations}

In order to make predictions about the physical properties of the thermal
string, observable quantities must be defined. At equilibrium, the
thermodynamical functions and their derivatives provide such observables. In
the nonequilibrium, it is more natural to express the observables as
transition probabilities for string states at different values of the time
parameter. These transition probabilities can be defined in terms of
correlations functions. Since we are considering the string at zero order in
the Euler number, i. e. on cylinder, and in the $a-b$ representation there are
no interactions among string modes, one would naively expect that the
correlation numbers be defined by%
\begin{equation}
\left\langle T[\phi^{i_{1}\alpha_{1}}(\tau_{1})\phi^{i_{2}\alpha_{2}}(\tau
_{2})\cdots\bar{\phi}^{i_{n}\alpha_{n}}(\tau_{n})]\right\rangle
.\label{corr-1}%
\end{equation}
However, the analysis from the section shows that even without interactions,
there is a counterterm $\hat{Q}$ that is generated in the $\hat{H}_{osc}$ as a
consequence of the the time evolution of the frequencies. In the interacting
field theory, $\hat{Q}$\ should be cancelled by self-energy interaction in
order to leave the Hamiltonian unchanged \cite{hu}. Therefore, in order to
define the correlation functions one has to work in the interaction picture
defined by the interaction Hamiltonian%
\begin{equation}
\hat{H}_{I}=\hat{Q}.\label{int-pict}%
\end{equation}
In this representation the oscillators evolve according as given in the
equations (\ref{eq-mot-therm-1}) and (\ref{eq-mot-therm-2}), that is with
respect to the free Hamiltonian $\hat{H}_{Q}$ as defined by the relation
(\ref{H-total-int}). Then the correlation functions are defined as the
expectation values%
\begin{equation}
\left\langle T[\phi^{i_{1}\alpha_{1}}(\tau_{1})\phi^{i_{2}\alpha_{2}}(\tau
_{2})\cdots\bar{\phi}^{i_{n}\alpha_{n}}(\tau_{n})\hat{S}(0,\infty
)]\right\rangle ,\label{corr-2}%
\end{equation}
where the $\hat{S}$ - operator is defined as usual
\begin{equation}
\hat{S}(0,\infty)=\lim_{\tau_{1}\rightarrow0}\lim_{\tau_{2}\rightarrow\infty
}\hat{u}(\tau_{1},\tau_{2})=T\left[  \exp\left(  -i\int_{0}^{\infty}%
d\varsigma\hat{H}_{I}(\varsigma)\right)  \right]  .\label{S-operator}%
\end{equation}
The thermal vacuum in the interaction picture satisfies the following relation%
\begin{equation}
\left\langle 0\right\vert \hat{u}(0,\tau)=\left\langle 0\right\vert
.\label{u-operator}%
\end{equation}
This is the Schwinger-Dyson formalism for the thermal fields in the canonical
formulation. The correlations defined by the relation (\ref{corr-2}) can be
calculated by expanding in powers of $\hat{Q}$%
\begin{align}
& \left\langle T[\phi^{i_{1}\alpha_{1}}(\tau_{1})\phi^{i_{2}\alpha_{2}}%
(\tau_{2})\cdots\bar{\phi}^{i_{n}\alpha_{n}}(\tau_{n})\hat{S}(0,\infty
)]\right\rangle =\nonumber\\
& \left\langle T[\phi^{i_{1}\alpha_{1}}(\tau_{1})\phi^{i_{2}\alpha_{2}}%
(\tau_{2})\cdots\bar{\phi}^{i_{n}\alpha_{n}}(\tau_{n})\right\rangle
-i\left\langle T[\phi^{i_{1}\alpha_{1}}(\tau_{1})\phi^{i_{2}\alpha_{2}}%
(\tau_{2})\cdots\bar{\phi}^{i_{n}\alpha_{n}}(\tau_{n})\int_{0}^{\infty
}d\varsigma\hat{H}_{I}(\varsigma)]\right\rangle +\cdots.\label{exp-corr}%
\end{align}
Note that the expansion is valid for small values of the time derivatives of
$n_{|\nabla|}$. Thus, the NETFD\ gives the general method to compute the
correlations of the thermal string in a perturbative like fashion.

In order to exemplify this formalism, let us compute the two points correlator
at zero order. To this end we consider the first term from (\ref{exp-corr}) in
the time-independent vacuum%
\begin{equation}
D^{ij,\alpha\beta}\left(  \tau_{1},\tau_{2};\sigma_{1},\sigma_{2}\right)
=-i\left\langle T\left[  \phi^{i\alpha}(\tau_{1},\sigma_{1})\bar{\phi}%
^{j\beta}(\tau_{2},\sigma_{2})\right]  \right\rangle . \label{def-therm-prop}%
\end{equation}
The fields from the equation (\ref{free-sol-eq-mot-bar}) act on the direct
space $\widehat{\mathcal{H}}$ without mixing the $a$ and $b$ sectors. It
follows that the Fourier transform of the oscillator propagators has the same
form in the two sectors and the cross terms vanish. Thus, there are just three
non-vanishing terms in the left-hand side of the relation
(\ref{def-therm-prop}) that correspond to the three distinct Hilbert spaces of
the string fields%
\begin{equation}
D^{ij,\alpha\beta}\left(  \tau_{1},\tau_{2};\sigma_{1},\sigma_{2}\right)
=D_{0}^{ij,\alpha\beta}\left(  \tau_{1},\tau_{2}\right)  +D_{a}^{ij,\alpha
\beta}\left(  \tau_{1},\tau_{2};\sigma_{1},\sigma_{2}\right)  +D_{b}%
^{ij,\alpha\beta}\left(  \tau_{1},\tau_{2};\sigma_{1},\sigma_{2}\right)  ,
\label{thermal-prop-sum}%
\end{equation}
where lower indices denote the center of mass and the corresponding sectors of
the string oscillators. By substituting the zero modes from (\ref{coord-cm-2})
into the above relation, one can show that the propagator of the center of
mass has the following form%
\begin{align}
D_{0}^{ij,\alpha\beta}\left(  \tau_{1},\tau_{2}\right)  =  &  -\frac{i}%
{2}\theta\left(  \tau_{1}-\tau_{2}\right)  \left[  U(\tau_{1})U(\tau
_{2})+U(\tau_{1})U^{\ast}(\tau_{2})\right]  \left(  \mathbf{1}_{2}%
+s_{3}\right)  ^{\alpha\beta}\delta^{ij}\nonumber\\
&  -\frac{i}{2}\theta\left(  \tau_{2}-\tau_{1}\right)  \left[  -U(\tau
_{2})U(\tau_{1})+U^{\ast}(\tau_{2})U(\tau_{1})\right]  \left(  \mathbf{1}%
_{2}-s_{3}\right)  ^{\alpha\beta}\delta^{ij}. \label{prop-cm}%
\end{align}

The calculation of the propagators of the oscillator sectors $a$ and $b$ from
the equation (\ref{thermal-prop-sum})\ is somewhat lengthy but not too
difficult. Since the sectors $a$ and $b$ can be related by the reflection
transformation $\sigma_{1}-\sigma_{2}\rightarrow\sigma_{2}-\sigma_{1}$ one can
calculate $D_{a}^{ij,\alpha\beta}\left(  \tau_{1},\tau_{2};\sigma_{1}%
,\sigma_{2}\right)  $ and obtain $D_{b}^{ij,\alpha\beta}\left(  \tau_{1}%
,\tau_{2};\sigma_{1},\sigma_{2}\right)  $ from it by applying the reflection.
In order to compute the expectation value from $D_{a}^{ij,\alpha\beta}\left(
\tau_{1},\tau_{2};\sigma_{1},\sigma_{2}\right)  $ we use the Fourier
decompositions (\ref{free-sol-eq-mot-bar}) and (\ref{free-sol-eq-mot-bar-bar})
and apply the Bogoliubov transformations (\ref{Bog-op-a}) and (\ref{Bog-op-b})
to transform the operators $a\left(  \tau\right)  $ into the time-independent
quasi-particle operators $\xi$ that correspond to the vacuum from the equation
(\ref{def-therm-prop}). After doing the algebra, the result takes the
following form%
\begin{align}
D_{a}^{ij,\alpha\beta}\left(  \tau_{1},\tau_{2};\sigma_{1},\sigma_{2}\right)
= &  -\frac{i\alpha^{\prime}}{4}\delta^{ij}\sum_{n>0}\left\{  B_{n}^{-1}%
(\tau_{1})\left[  e^{2in(\sigma_{1}-\sigma_{2})}D_{a,+}^{ij}\left(  \tau
_{1},\tau_{2};n\right)  \right.  \right.  \nonumber\\
&  \left.  \left.  +e^{-2in(\sigma_{1}-\sigma_{2})}D_{a,-}^{ij}\left(
\tau_{1},\tau_{2};n\right)  \right]  B_{n}(\tau_{2})\right\}  ^{\alpha\beta
}.\label{prop-a}%
\end{align}
We have denoted by the indices $+$ and $-$ terms that would correspond to the
advanced and retarded propagators, respectively, if the frequency did not
depend on time. These terms are given by the following equations%
\begin{align}
D_{a,+}^{\alpha\beta}\left(  \tau_{1},\tau_{2};n\right)  = &  \frac
{\exp\left[  i\int_{\tau_{1}}^{\tau_{2}}d\lambda\omega_{n}(\lambda)\right]
}{2n\sqrt{\omega_{n}\left(  \tau_{1}\right)  \omega_{n}\left(  \tau
_{2}\right)  }}\left[  \theta\left(  \tau_{1}-\tau_{2}\right)  \left(
\mathbf{1}_{2}+s_{3}\right)  +\theta\left(  \tau_{2}-\tau_{1}\right)  \left(
\mathbf{1}_{2}-s_{3}\right)  \right]  ^{\alpha\beta},\label{prop-a-adv}\\
D_{a,-}^{\alpha\beta}\left(  \tau_{1},\tau_{2};n\right)  = &  \frac
{\exp\left[  -i\int_{\tau_{1}}^{\tau_{2}}d\lambda\omega_{n}(\lambda)\right]
}{2n\sqrt{\omega_{n}\left(  \tau_{1}\right)  \omega_{n}\left(  \tau
_{2}\right)  }}\left[  C_{n}(\tau_{1})\left[  \theta\left(  \tau_{1}-\tau
_{2}\right)  \left(  \mathbf{1}_{2}+s_{3}\right)  \right.  \right.
\nonumber\\
&  \left.  \left.  +\theta\left(  \tau_{2}-\tau_{1}\right)  \left(
\mathbf{1}_{2}-s_{3}\right)  \right]  C_{n}^{-1}(\tau_{2})\right]
^{\alpha\beta},\label{prop-a-ret}%
\end{align}
where we have introduced the matrix for each mode $m$%
\begin{equation}
C_{m}(\tau)=%
\begin{pmatrix}
1+2n_{m}(\tau) & -1\\
-1 & 0
\end{pmatrix}
.\label{C-matrix}%
\end{equation}
The Fourier transforms of the nonequilibrium propagators in the $a$ sector
given by the relations (\ref{prop-a-adv}) and (\ref{prop-a-ret}) are similar
to the nonequilibrium propagators of the scalar field in the Minkowski
spacetime \cite{mi,miny,mi2}. As a difference from the scalar field, we note
the factor $\sqrt[-1]{\omega_{n}\left(  \tau_{1}\right)  \omega_{n}\left(
\tau_{2}\right)  }$ that is not present in the Minkowski spacetime. This
factor is a consequence of the dependence of the field modes $\phi
_{n}^{i\alpha}(\tau_{1})$ on $\sqrt[-1]{n\omega_{n}\left(  \tau\right)  }$,
more specifically,it is the result of the interaction between the
gravitational background and the strings. The nonequilibrium propagators from
the $b$ sector can be calculated in exactly the same way.

\section{Conclusions}

In this paper, we have given a canonical method to study the strings near
singularity of plane wave backgrounds based on the NETFD. This method allows
one to construct and interpret the Hilbert space and to compute the
correlation functions of thermal string. The correlation functions of the
string oscillators receive corrections from the counterterm $\hat{Q}$ that
generates the time evolution of the number parameter $n_{|\nabla|}(\tau)$.
Compared with the literature, these results are similar to the ones that are
obtained in the case of relativistic scalar field in Minkowski
\cite{mi,miny,mi2}. However, the extra degrees of freedom provided by the
center of mass of string do not have any analogue in the thermal field theory.
Therefore, in order to compute their thermal behavior, we had to generalize
the NETFD by postulating a new commutation relation for the canonically
conjugate variables of the center of mass (\ref{cm-comm}) that is consistent
with the equal time commutators of the thermal fields given by the relations
(\ref{eq-t-comm-1}), (\ref{eq-t-comm-2}) and (\ref{eq-t-comm-3}). In the limit
of small time variations of the number parameter, the NETFD provides a
perturbative method to compute the correlation function that we have used to
determine the two point functions at zero order. Due to the importance of the
applications of the nonequilibrium string theory, the development of the
method presented in this paper deserves attention. In particular, it is
important to calculate the two point functions at first order and the equation
that describe the evolution of the thermal correlators. We will report on
these topics elsewhere.

\noindent\textbf{Acknowledgements} I. V. V. would like to thank to D.
Berenstein for discussions and to the organizers of LASS 2010 where this work
began. Also, I. V. V. would like to thank to J. A. Helay\"{e}l-Neto and A. M.
O. de Almeida for hospitality at LAFEX-CBPF where part of this work was
accomplished. R. N. acknowledges a FAPERJ fellowship.

\newpage\begin{appendix}
\section{Nonequilibrium Thermo Field Dynamics}
\end{appendix}

The NETFD is a real-time canonical formalism of thermal quantum field theory.
It has been originally developed as an alternative to the nonequilibrium
imaginary-time formalisms which presented difficulties in defining the
multi-time correlators and the density operators \cite{hu}.The NETFD can be
derived from the same basic set of axioms as the equilibrium TFD which can be
summarized as follows

\begin{enumerate}
\item The thermal physical system is described by two sets of commuting
(anti-commuting) field operators $\phi\left(  x\right)  $ and $\tilde{\phi
}\left(  x\right)  $, respectively, that act on the thermal or total Hilbert
space which is the direct product of two identical Hilbert spaces
\begin{equation}
\widehat{\mathcal{H}}=\mathcal{H}\otimes\mathcal{\tilde{H}}. \label{A-1}%
\end{equation}

\item To operators $O=O\left(  \phi\left(  x\right)  ,\phi^{\dag}\left(
x\right)  \right)  $ that act on $\widehat{\mathcal{H}}$ are associated
operators $\tilde{O}=O^{\ast}\left(  \tilde{\phi}\left(  x\right)
,\tilde{\phi}^{\dag}\left(  x\right)  \right)  $.

\item The tilde defines an involution on $\widehat{\mathcal{H}}$ that obeys
the following rules%
\begin{align}
\left(  c_{1}O_{1}+c_{2}O_{2}\right)  ^{\symbol{126}}  &  =c_{1}^{\ast}%
\tilde{O}_{1}+c_{2}^{\ast}\tilde{O}_{2},\label{A2-1}\\
\left(  O_{1}O_{2}\right)  ^{\symbol{126}}  &  =\tilde{O}_{1}\tilde{O}%
_{2},\label{A2-2}\\
\left(  O^{\dag}\right)  ^{\symbol{126}}  &  =\tilde{O}^{\dag},\label{A2-3}\\
\left(  \tilde{O}\right)  ^{\symbol{126}}  &  =\varepsilon O, \label{A2-4}%
\end{align}
for all $c_{1},c_{2}\in%
%TCIMACRO{\U{2102} }%
%BeginExpansion
\mathbb{C}
%EndExpansion
$ and all $O,O_{1},O_{2}\in End(\widehat{\mathcal{H}})$. $\varepsilon=+1(-1)$
for bosonic (fermionic) operators.

\item The vacuum state is invariant under the involution%
\begin{equation}
\widetilde{\left\vert 0(t)\right\rangle }=\left\vert 0(t)\right\rangle
,\qquad\widetilde{\left\langle 0(t)\right\vert }=\left\langle 0(t)\right\vert
. \label{A-3}%
\end{equation}

\item The time evolution is generator $\hat{H}$ should satisfy the following
condition%
\begin{equation}
\left(  i\hat{H}\right)  ^{\symbol{126}}=i\hat{H}. \label{A-4}%
\end{equation}
From this condition one can see that the Hamiltonian is the difference between
the two identical Hamiltonians
\begin{equation}
\hat{H}=H-\tilde{H} \label{A-4-a}%
\end{equation}
and it is given by the Heisenberg equation for non-tilde operators%
\begin{equation}
i\frac{d}{dt}O(t)=\left[  O,\hat{H}\right]  . \label{A-5}%
\end{equation}

\item There is a set of operators $\left\{  \xi,\tilde{\xi},\xi^{\dag}%
,\tilde{\xi}^{\dag}\right\}  $that define the time-independent free
quasi-particle representation and have the following action on the vacuum%
\begin{equation}
\xi\left\vert 0\right\rangle =\tilde{\xi}\left\vert 0\right\rangle
=0,\qquad\left\langle 0\right\vert \xi^{\dag}=\left\langle 0\right\vert
\tilde{\xi}^{\dag}=0. \label{A-6}%
\end{equation}
The thermal states are defined by the following \emph{thermal state condition}%
\begin{align}
\left\langle 0\right\vert O(t)  &  =\left\langle 0\right\vert \tilde{O}^{\dag
}(t)\qquad\text{for bosons},\label{A-6-a}\\
\left\langle 0\right\vert O(t)  &  =e^{i\theta}\left\langle 0\right\vert
\tilde{O}^{\dag}(t)\qquad\text{for fermions}, \label{A-6-b}%
\end{align}
where $\theta$ is determined by the tilde conjugation rules.

\item The thermal average of the dynamical observable $O$ is given by%
\begin{equation}
\left\langle O\right\rangle =\left\langle 0\left\vert O\right\vert
0\right\rangle . \label{A-7}%
\end{equation}

\item For any value of the time variable $t$, there is an invertible map
between the total Hilbert space and the time-dependent quasi-particle
representation given by a time-dependent Bogoliubov transformation%
\begin{equation}%
\begin{pmatrix}
\phi\left(  t\right) \\
e^{i\theta}\tilde{\phi}\left(  t\right)
\end{pmatrix}
=B^{-1}\left(  t\right)
\begin{pmatrix}
\xi\left(  t\right) \\
e^{i\theta}\tilde{\xi}\left(  t\right)
\end{pmatrix}
. \label{A-8}%
\end{equation}

\item The ket-vacuum and the bra-vacuum of the time-dependent representation
is defined as%
\begin{align}
\xi\left(  t\right)  \left\vert 0(t)\right\rangle  &  =\tilde{\xi}\left(
t\right)  \left\vert 0(t)\right\rangle =0,\label{A-9-a1}\\
\left\langle 0(t)\right\vert \xi^{\dag}\left(  t\right)   &  =\left\langle
0(t)\right\vert \tilde{\xi}^{\dag}\left(  t\right)  =0. \label{A-9-b1}%
\end{align}
The time-evolution of the time-dependent bra- and ket-vacua is given by the
equations%
\begin{align}
i\frac{\partial}{\partial t}\left\vert 0(t)\right\rangle  &  =\hat
{H}\left\vert 0(t)\right\rangle ,\label{A-9-a2}\\
\left\langle 0(t)\right\vert \hat{H}  &  =0. \label{A-9-b2}%
\end{align}

\item The stationary thermal states in the Schroedinger representation are
defined by the equation%
\begin{equation}
\lim_{t\rightarrow\infty}\hat{H}\left\vert 0(t)\right\rangle =0,\label{A-10}%
\end{equation}
assuming that the thermal equilibrium is obtained at $t\rightarrow\infty$.
\end{enumerate}

The NETFD is a canonical formalism. Therefore, there is the time-dependent
canonical representation defined in terms of the canonical operators $\left\{
a(t),a^{\dagger}(t),\tilde{a}(t),\tilde{a}^{\dagger}(t)\right\}  $. The family
of time-parametrized mappings between the canonical representation and any
time-dependent quasi-particle representation $\left\{  \xi\left(  t\right)
,\xi^{\dag}\left(  t\right)  ,\tilde{\xi}\left(  t\right)  ,\tilde{\xi}^{\dag
}\left(  t\right)  \right\}  $ is given by the time-dependent invertible
Bogoliubov map%
\begin{equation}
B(t):\left\{  a(t),a^{\dagger}(t),\tilde{a}(t),\tilde{a}^{\dagger}(t)\right\}
\longrightarrow\left\{  \xi\left(  t\right)  ,\xi^{\dag}\left(  t\right)
,\tilde{\xi}\left(  t\right)  ,\tilde{\xi}^{\dag}\left(  t\right)  \right\}  .
\label{A-11}%
\end{equation}
The quasi-particle representation is defined by the properties of the
Bogoliubov map that can satisfy three conditions: i) it can preserve the
canonical structure; ii) it can preserve the Hermitian conjugation and iii) it
can preserve the tilde conjugation:%
\begin{align}
B(t)\left(  s_{2}\otimes s_{3}\right)  B^{T}(t)  &  =s_{2}\otimes
s_{3},\label{A-12-a}\\
B^{\ast}(t)\left(  s_{1}\otimes\mathbf{1}\right)   &  =\left(  s_{1}%
\otimes\mathbf{1}\right)  B(t),\label{A-12-b}\\
B^{\ast}(t)\left(  s_{1}\otimes s_{1}\right)   &  =\left(  s_{1}\otimes
s_{1}\right)  B(t), \label{A-12-c}%
\end{align}
where $s_{1}$, $s_{2}$ and $s_{3}$ are the Pauli matrices. The first
requirement is made in all cases. The last two ones define non-standard,
albeit useful representations of the thermal field theory. It can be shown
that the most general form of the non-Hermitian Bogoliubov operator in the
doublet representation from equation (\ref{A-8}) is%
\begin{equation}
B_{\mathbf{k}}(t)=\left(  1+\varepsilon n_{\mathbf{k}}\left(  t\right)
\right)  ^{\frac{1}{2}}e^{\gamma_{\mathbf{k}}(t)s_{3}}%
\begin{pmatrix}
1 & -f_{\mathbf{k}}^{\alpha_{k}}(t)\\
-\varepsilon f_{\mathbf{k}}^{1-\alpha_{\mathbf{k}}}(t) & 1
\end{pmatrix}
\theta_{k}\left(  t\right)  , \label{A-13}%
\end{equation}
where $\mathbf{k}$ denotes the canonical mode of the field, $\varepsilon
=+1(-1)$ for bosons (fermions) and $\alpha_{\mathbf{k}}$, $n_{\mathbf{k}}(t)$
and $\gamma_{\mathbf{k}}(t)$ are parameters of the gauge group of thermal
mappings which is $SU(1,1)$ for bosons and $SO(2)$ for fermions. The pure
complex phase function $\theta_{k}\left(  t\right)  $ depends on the energy of
the mode and satisfies the first order differential equation%
\begin{equation}
i\frac{d}{d\tau}\theta_{n}\left(  \tau\right)  =\omega_{n}(\tau)\theta
_{n}\left(  \tau\right)  . \label{A-13-a}%
\end{equation}
The physical interpretation $n_{\mathbf{k}}(t)$ is that of the number density
defined as%
\begin{equation}
n_{\mathbf{k}}(t)\delta\left(  \mathbf{k-l}\right)  =\left\langle
0(t)\left\vert a_{\mathbf{k}}^{\dag}a_{\mathbf{l}}\right\vert
0(t)\right\rangle . \label{A-14}%
\end{equation}
The function $f_{\mathbf{k}}^{\alpha_{k}}(t)$ is the statistical distribution%
\begin{equation}
f_{\mathbf{k}}^{\alpha_{k}}(t)=\frac{n_{\mathbf{k}}(t)}{1+\varepsilon
n_{\mathbf{k}}(t)}. \label{A-15}%
\end{equation}
The parameters $\alpha_{\mathbf{k}}\in\left[  0,1\right]  $ lead, in general,
to different representations of the thermal field in the Liouville space. The
choice of the gauge $\alpha_{\mathbf{k}}=1$ and $\gamma_{\mathbf{k}}%
(t)=\ln\left[  1+\varepsilon n_{\mathbf{k}}(t)\right]  $ renders the
Bogoliubov transformations linear (\cite{ph})%
\begin{equation}
B_{\mathbf{k}}(t)=%
\begin{pmatrix}
1+\varepsilon n_{\mathbf{k}}(t) & -n_{\mathbf{k}}(t)\\
-\varepsilon & 1
\end{pmatrix}
. \label{A-16}%
\end{equation}


\newpage

\begin{thebibliography}{99}                                                                                               %
%\cite{Blau:2002dy}


%\cite{deVega:1995bq}


\bibitem {vs} H.~J.~de Vega and N.~G.~Sanchez,
%``Lectures on string theory in curved space-times,''
In *Erice 1995, String gravity and physics at the Planck energy scale* 11-63
[hep-th/9512074].
%%CITATION = HEP-TH/9512074;%%


\bibitem {bfhp}M.~Blau, J.~M.~Figueroa-O'Farrill, C.~Hull and
G.~Papadopoulos,
%``Penrose limits and maximal supersymmetry,''
Class.\ Quant.\ Grav.\ \textbf{19}, L87 (2002) [arXiv:hep-th/0201081].
%%CITATION = CQGRD,19,L87;%%


%\cite{Russo:2002qj}


\bibitem {rt}J.~G.~Russo and A.~A.~Tseytlin,
%``A Class of exact pp wave string models with interacting light cone gauge
%actions,''
JHEP \textbf{0209}, 035 (2002) [arXiv:hep-th/0208114].
%%CITATION = JHEPA,0209,035;%%


%\cite{Papadopoulos:2002bg}


\bibitem {prt}G.~Papadopoulos, J.~G.~Russo, A.~A.~Tseytlin,
%``Solvable model of strings in a time dependent plane wave background,''
Class.\ Quant.\ Grav.\ \textbf{20}, 969-1016 (2003). [hep-th/0211289].

%\cite{Craps:2008bv}


\bibitem {cre}B.~Craps, F.~De Roo and O.~Evnin,
%``Can free strings propagate across plane wave singularities?,''
JHEP \textbf{0903}, 105 (2009) [arXiv:0812.2900 [hep-th]].
%%CITATION = JHEPA,0903,105;%%


%\cite{Craps:2002ii}


\bibitem {ckr}B.~Craps, D.~Kutasov and G.~Rajesh,
%``String propagation in the presence of cosmological singularities,''
JHEP \textbf{0206}, 053 (2002) [arXiv:hep-th/0205101].
%%CITATION = JHEPA,0206,053;%%


%\cite{Narayan:2009pu}


\bibitem {kn1}K.~Narayan,
%``Null cosmological singularities and free strings,''
Phys.\ Rev.\ D \textbf{81}, 066005 (2010) [arXiv:0909.4731 [hep-th]].
%%CITATION = PHRVA,D81,066005;%%


%\cite{Narayan:2010rm}


\bibitem {kn2}K.~Narayan,
%``Null cosmological singularities and free strings: II,''
JHEP \textbf{1101}, 145 (2011) [arXiv:1012.0113 [hep-th]].
%%CITATION = JHEPA,1101,145;%%


%\cite{Craps:2011sp}


\bibitem {ce}B.~Craps, O.~Evnin,
%``Light-like Big Bang singularities in string and matrix theories,''
[arXiv:1103.5911 [hep-th]].

%\cite{Gadelha:2006dm}


\bibitem {gmn}A.~L.~Gadelha, D.~Z.~Marchioro, D.~L.~Nedel,
%``Entanglement and entropy operator for strings in pp-wave time dependent background,''
Phys.\ Lett.\ \textbf{B639}, 383-388 (2006). [hep-th/0605237].

%-------- STRINGS AT THERMAL EQUILIBRIUM HAGEDORN ---------------------------------------------------


%\cite{PandoZayas:2002hh}


\bibitem {zv}L.~A.~Pando Zayas and D.~Vaman,
%``Strings in RR plane wave background at finite temperature,''
Phys.\ Rev.\ D \textbf{67}, 106006 (2003) [arXiv:hep-th/0208066].
%%CITATION = PHRVA,D67,106006;%%


%\cite{Greene:2002cd}


\bibitem {gss}B.~R.~Greene, K.~Schalm and G.~Shiu,
%``On the Hagedorn behaviour of PP wave strings and N=4 SYM theory at finite R
%charge density,''
Nucl.\ Phys.\ B \textbf{652}, 105 (2003) [arXiv:hep-th/0208163].
%%CITATION = NUPHA,B652,105;%%


%\cite{Sugawara:2002rs}


\bibitem {ys1}Y.~Sugawara,
%``Thermal amplitudes in DLCQ superstrings on PP waves,''
Nucl.\ Phys.\ B \textbf{650}, 75 (2003) [arXiv:hep-th/0209145].
%%CITATION = NUPHA,B650,75;%%


%\cite{Sugawara:2003qc}


\bibitem {ys2}Y.~Sugawara,
%``Thermal partition function of superstring on compactified PP wave,''
Nucl.\ Phys.\ B \textbf{661}, 191 (2003) [arXiv:hep-th/0301035].
%%CITATION = NUPHA,B661,191;%%


%\cite{Brower:2002zx}


\bibitem {blt}R.~C.~Brower, D.~A.~Lowe and C.~I.~Tan,
%``Hagedorn transition for strings on pp waves and tori with chemical
%potentials,''
Nucl.\ Phys.\ B \textbf{652}, 127 (2003) [arXiv:hep-th/0211201].
%%CITATION = NUPHA,B652,127;%%


%\cite{Grignani:2003cs}


\bibitem {gost}G.~Grignani, M.~Orselli, G.~W.~Semenoff and D.~Trancanelli,
%``The Superstring Hagedorn temperature in a pp wave background,''
JHEP \textbf{0306}, 006 (2003) [arXiv:hep-th/0301186].
%%CITATION = JHEPA,0306,006;%%


%\cite{Hyun:2003ks}


\bibitem {hpy}S.~j.~Hyun, J.~D.~Park and S.~H.~Yi,
%``Thermodynamic behavior of IIA string theory on a pp wave,''
JHEP \textbf{0311}, 006 (2003) [arXiv:hep-th/0304239].
%%CITATION = JHEPA,0311,006;%%


%\cite{Hashimoto:2004ve}


\bibitem {hz}A.~Hashimoto and L.~Pando Zayas,
%``Correspondence principle for black holes in plane waves,''
JHEP \textbf{0403}, 014 (2004) [arXiv:hep-th/0401197].
%%CITATION = JHEPA,0403,014;%%


%\cite{Bigazzi:2003jk}


\bibitem {bc}F.~Bigazzi and A.~L.~Cotrone,
%``On zero point energy, stability and Hagedorn behavior of type IIB strings
%on pp waves,''
JHEP \textbf{0308}, 052 (2003) [arXiv:hep-th/0306102].
%%CITATION = JHEPA,0308,052;%%


%----------------------------------------------------------------------------------------------------


%\cite{Evnin:2008ya}


\bibitem {en}O.~Evnin and T.~Nguyen,
%``On discrete features of the wave equation in singular pp-wave
%backgrounds,''
JHEP \textbf{0809}, 105 (2008) [arXiv:0806.3057 [hep-th]].
%%CITATION = JHEPA,0809,105;%%


%\cite{Kim:1997uk}


\bibitem {sk}S.~P.~Kim, ``Liouville-Neumann approach to the nonperturbative
quantum field theory,'' Proceedings "Seoul 1997, Recent developments in
nonperturbative quantum field theory", page. 345-352, arXiv:hep-th/9706052.
%%CITATION = HEP-TH/9706052;%%


%\cite{Kim:2000xb}


\bibitem {kl}S.~P.~Kim, C.~H.~Lee,
%``Nonequilibrium quantum dynamics of second order phase transitions,''
Phys.\ Rev.\ \textbf{D62}, 125020 (2000). [hep-ph/0005224].

%\cite{Lewis:1968tm}


\bibitem {lr}H.~R.~Lewis, W.~B.~Riesenfeld,
%``An Exact quantum theory of the time dependent harmonic oscillator and of a charged particle time dependent electromagnetic field,''
J.\ Math.\ Phys.\ \textbf{10}, 1458-1473 (1969).

%---------------RESULTS IN STRING OBTAINED WITH LvN METHOD---------------------------------


%\cite{Gimon:2002sf}


\bibitem {gzs}E.~G.~Gimon, L.~A.~Pando Zayas, J.~Sonnenschein,
%``Penrose limits and RG flows,''
JHEP \textbf{0209}, 044 (2002). [hep-th/0206033].

%\cite{Sfetsos:2003vw}


\bibitem {ks}K.~Sfetsos,
%``The Exact description of NS5-branes in the Penrose limit,''
Nucl.\ Phys.\ \textbf{B669}, 103-127 (2003). [hep-th/0305109].

%\cite{Blau:2004cm}


\bibitem {bbl}M.~Blau, M.~Borunda, M.~O'Loughlin,
%``On the Hagedorn behaviour of singular scale-invariant plane waves,''
JHEP \textbf{0510}, 047 (2005). [hep-th/0412228].

%-------------------------------------------------------------------------------------


%\cite{Umezawa:1993yq}


\bibitem {hu}H.~Umezawa, ``Advanced field theory: Micro, macro, and thermal
physics,'' New York, USA: AIP (1993).
%238 p.


%---------- OLD TFD and STRINGS ----------------------------------------


%\cite{Leblanc:1987zj}


\bibitem {yl1}Y.~Leblanc,
%``String Field Theory At Finite Temperature,''
Phys.\ Rev.\ D \textbf{36}, 1780 (1987).
%%CITATION = PHRVA,D36,1780;%%


%\cite{Leblanc:1987hw}


\bibitem {yl2}Y.~Leblanc,
%``Finite Temperature Amplitudes In Open String Systems,''
Phys.\ Rev.\ D \textbf{37} 1547 (1988).
%%CITATION = PHRVA,D37,1547;%%


%\cite{Leblanc:1988eq}


\bibitem {yl3}Y.~Leblanc,
%``Cosmological Aspects Of The Heterotic String Above The Hagedorn
%Temperature,''
Phys.\ Rev.\ D \textbf{38}, 3087 (1988).
%%CITATION = PHRVA,D38,3087;%%


%\cite{Leblanc:1989gs}


\bibitem {yl4}Y.~Leblanc,
%``Improved Integral Representation For The Finite-Temperature Propagator In
%String Theory,''
Phys.\ Rev.\ D \textbf{39} 1139 (1989).
%%CITATION = PHRVA,D39,1139;%%


%\cite{Leblanc:1990ub}


\bibitem {yl5}Y.~Leblanc, M.~Knecht and J.~C.~Wallet,
%``Regularization Of Finite Temperature String Theories,''
Phys.\ Lett.\ B \textbf{237} 357 (1990).
%%CITATION = PHLTA,B237,357;%%


%\cite{Leblanc:1990hx}


\bibitem {yl6}Y.~Leblanc,
%``Generalized Mcclain-Roth-O'brien-Tan Theorem And The String Free Energy At
%Higher Genus,''
Phys.\ Rev.\ Lett.\ \textbf{64} 831 (1990).
%%CITATION = PRLTA,64,831;%%


%\cite{Fujisaki:1988ci}


\bibitem {fns1}H.~Fujisaki, K.~Nakagawa and I.~Shirai,
%``Comments On The Thermal Neveu-Schwarz-Ramond Superstring,''
Prog.\ Theor.\ Phys.\ \textbf{81}, 570 (1989).
%%CITATION = PTPKA,81,570;%%


%\cite{Fujisaki:1989kp}


\bibitem {fn1}H.~Fujisaki and K.~Nakagawa,
%``The Thermal Stability Of Renormalization Of Open Bosonic Strings,''
Prog.\ Theor.\ Phys.\ \textbf{82}, 236 (1989).
%%CITATION = PTPKA,82,236;%%


%\cite{Fujisaki:1989sy}


\bibitem {fn2}H.~Fujisaki and K.~Nakagawa,
%``Renormalizability Of Open Bosonic Thermal Strings,''
Prog.\ Theor.\ Phys.\ \textbf{82}, 1017 (1989).
%%CITATION = PTPKA,82,1017;%%


%\cite{Fujisaki:1990vw}


\bibitem {fn3}H.~Fujisaki and K.~Nakagawa,
%``Infrared Behavior Of The Type I Neveu-Schwarz-Ramond Open Thermal
%Superstring,''
Europhys.\ Lett.\ \textbf{14}, 737 (1991).
%%CITATION = EULEE,14,737;%%


%\cite{Fujisaki:1992sw}


\bibitem {fn4}H.~Fujisaki and K.~Nakagawa,
%``Infrared behavior of the type II closed thermal superstring,''
Europhys.\ Lett.\ \textbf{20}, 677 (1992).
%%CITATION = EULEE,20,677;%%


%\cite{Fujisaki:1993bx}


\bibitem {fn5}H.~Fujisaki,
%``Dimensional regularization of the closed bosonic thermal string,''
Europhys.\ Lett.\ \textbf{28}, 623 (1994).
%%CITATION = EULEE,28,623;%%


%\cite{Fujisaki:1997rc}


\bibitem {f1}H.~Fujisaki,
%``Thermofield dynamics of the heterotic string: Physical aspects of the
%thermal duality,''
Europhys.\ Lett.\ \textbf{39}, 479 (1997).
%[arXiv:hep-th/9704180].
%%CITATION = HEP-TH 9704180;%%


%----- GADELHA-NEDEL ----------------------------------------------------


%\cite{Nedel:2004gy}


\bibitem {ng1}D.~L.~Nedel, M.~C.~B.~Abdalla and A.~L.~Gadelha,
%``Superstring in a pp-wave background at finite temperature: TFD  approach,''
Phys.\ Lett.\ B \textbf{598}, 121 (2004).
%[arXiv:hep-th/0405258].
%%CITATION = HEP-TH 0405258;%%


%\cite{Abdalla:2005qs}


\bibitem {ng2}M.~C.~B.~Abdalla, A.~L.~Gadelha and D.~L.~Nedel,
%``PP-wave light-cone free string field theory at finite temperature,''
JHEP \textbf{0510}, 063 (2005).
%[arXiv:hep-th/0508195].
%%CITATION = HEP-TH 0508195;%%


%\cite{Abdalla:2004dg}


\bibitem {ng3}M.~C.~B.~Abdalla, A.~L.~Gadelha and D.~L.~Nedel,
%``Closed string thermal torus from thermofield dynamics,''
Phys.\ Lett.\ B \textbf{613}, 213 (2005).
%[arXiv:hep-th/0410068].
%%CITATION = HEP-TH 0410068;%%


%\cite{Abdalla:2004xn}


\bibitem {ng4}M.~C.~B.~Abdalla, A.~L.~Gadelha and D.~L.~Nedel,
%``Perspectives of TFD on string theory,''
PoS \textbf{WC2004}, 020 (2004).
%[arXiv:hep-th/0412134].
%%CITATION = HEP-TH 0412134;%%


%\cite{Abdalla:2004xg}


\bibitem {ng5}M.~C.~B.~Abdalla, A.~L.~Gadelha and D.~L.~Nedel,
%``General unitary TFD formulation for superstrings,''
PoS \textbf{WC2004}, 032 (2004).
%[arXiv:hep-th/0412128].
%%CITATION = HEP-TH 0412128;%%


%------------- ASPECTS of TFD in ADS ----------------------------------


%\cite{Abdalla:2000bg}


\bibitem {ivv9}M.~C.~B.~Abdalla, A.~L.~Gadelha and I.~V.~Vancea,
%``On (b,c)-system at finite temperature in thermo field approach,''
Phys.\ Lett.\ A \textbf{273}, 235 (2000).
%[arXiv:hep-th/0003209].
%%CITATION = PHLTA,A273,235;%%


%\cite{Belich:2006wc}


\bibitem {ivv10}H.~Belich, E.~L.~Graca, M.~A.~Santos and I.~V.~Vancea,
%``Semiclassical thermal string in the black-hole AdS spacetime,''
JHEP \textbf{0702}, 037 (2007).
%[arXiv:hep-th/0610271].
%%CITATION = JHEPA,0702,037;%%


%---- THERMAL D-BRANES ----------------------------------------------------------


%\cite{Vancea:2000gr}


\bibitem {ivv1}I.~V.~Vancea,
%``Bosonic D-branes at finite temperature,''
Phys.\ Lett.\ B \textbf{487}, 175 (2000).
%[arXiv:hep-th/0006228].
%%CITATION = PHLTA,B487,175;%%


%\cite{Abdalla:2001ad}


\bibitem {ivv2}M.~C.~B.~Abdalla, A.~L.~Gadelha and I.~V.~Vancea,
%``Bosonic D-branes at finite temperature with an external field,''
Phys.\ Rev.\ D \textbf{64}, 086005 (2001).
%[arXiv:hep-th/0104068].
%%CITATION = PHRVA,D64,086005;%%


%\cite{Abdalla:2002sb}


\bibitem {ivv3}M.~C.~B.~Abdalla, A.~L.~Gadelha and I.~V.~Vancea,
%``On the SU(1,1) thermal group of bosonic strings and D-branes,''
Phys.\ Rev.\ D \textbf{66}, 065005 (2002).
%[arXiv:hep-th/0203222].
%%CITATION = PHRVA,D66,065005;%%


%\cite{Vancea:2006wf}


\bibitem {ivv4}I.~V.~Vancea,
%``Thermal D-brane boundary states from Green-Schwarz superstrings,''
Phys.\ Rev.\ D \textbf{74}, 086002 (2006).
%[arXiv:hep-th/0607167].
%%CITATION = PHRVA,D74,086002;%%


%\cite{Vancea:2007ux}


\bibitem {ivv5}I.~V.~Vancea,
%``Thermal $D$-brane boundary states from type IIB Green-Schwarz superstring
%in {\em pp}-wave background,''
Int.\ J.\ Mod.\ Phys.\ A \textbf{23}, 4485 (2008).
%[arXiv:0712.1569 [hep-th]].
%%CITATION = IMPAE,A23,4485;%%


%-------------- MARCELO BOTA ----------------------------------


%\cite{Cantcheff:2007nr}


\bibitem {mbc1}M.~B.~Cantcheff,
%``D-branes as coherent states in the open string channel,''
Eur.\ Phys.\ J.\ C \textbf{55}, 517 (2008).
%[arXiv:0710.3186 [hep-th]].
%%CITATION = EPHJA,C55,517;%%


%\cite{Cantcheff:2009yf}


\bibitem {mbc2}M.~B.~Cantcheff,
%``String Entanglement and D-branes as Pure States,''
Phys.\ Rev.\ D \textbf{80}, 046001 (2009).
%[arXiv:0906.3049 [hep-th]].
%%CITATION = PHRVA,D80,046001;%%


%-------- REVIEWS THERMAL D-BRANES


%\cite{Abdalla:2003ki}


\bibitem {ivv6}M.~C.~B.~Abdalla, A.~L.~Gadelha and I.~V.~Vancea,
%``TFD approach for bosonic strings and D(p)-branes,''
Int.\ J.\ Mod.\ Phys.\ A \textbf{18}, 2109 (2003).
%[arXiv:hep-th/0301249].
%%CITATION = IMPAE,A18,2109;%%


%\cite{Abdalla:2003dx}


\bibitem {ivv7}M.~C.~B.~Abdalla, A.~L.~Gadelha and I.~V.~Vancea,
%``D-branes at finite temperature in TFD,''
arXiv:hep-th/0308114.
%%CITATION = HEP-TH/0308114;%%


%\cite{Abdalla:2004xs}


\bibitem {ivv8}M.~C.~B.~Abdalla, A.~L.~Gadelha and I.~V.~Vancea,
%``Bosonic Dp-branes at finite temperature in TFD approach,''
Nucl.\ Phys.\ Proc.\ Suppl.\ \textbf{127} (2004) 92.
%%CITATION = NUPHZ,127,92;%%


%\cite{Nardi:2010xc}


\bibitem {ivv11}R.~Nardi, M.~A.~Santos and I.~V.~Vancea,
%``Thermal magnetized D-branes on $R^{1,p}\times T^{d-p-1}$ in the generalized
%Thermo Field Dynamics approach,''
J.\ Phys.\ A \textbf{44}, 235403 (2011) [arXiv:1011.0574 [hep-th]].
%%CITATION = JPAGB,A44,235403;%%


%------------------------- OLD RELATIVISTIC NETFD ----------------------------------------


%\cite{Matsumoto:1988gf}


\bibitem {hm}H.~Matsumoto,
%``Quasiparticle Field In Nonequilibrium Quantum Field Theory,''
Physica \textbf{A158}, 291-305 (1989).

%\cite{Henning:1995sm}


\bibitem {ph}P.~A.~Henning,
%``Thermo field dynamics for quantum fields with continuous mass spectrum,''
Phys.\ Rept.\ \textbf{253}, 235-380 (1995).

%------------------------- NETFD BASIC REFERENCES ---------------------------------------


%\cite{Arimitsu:1985ez}


\bibitem {au1}T.~Arimitsu, H.~Umezawa,
%``A General Formulation Of Nonequilibrium Thermo Field Dynamics,''
Prog.\ Theor.\ Phys.\ \textbf{74}, 429 (1985).

%\cite{Arimitsu:1985xm}


\bibitem {au2}T.~Arimitsu, H.~Umezawa,
%``Nonequilibrium Thermo Field Dynamics,''
Prog.\ Theor.\ Phys.\ \textbf{77}, 32 (1987).

%\cite{Arimitsu:1985xn}


\bibitem {au3}T.~Arimitsu, H.~Umezawa,
%``General Structure Of Nonequilibrium Thermo Field Dynamics,''
Prog.\ Theor.\ Phys.\ \textbf{77}, 53 (1987).

%\cite{Umezawa:1988ii}


\bibitem {uy}H.~Umezawa, Y.~Yamanaka,
%``Selfconsistent Renormalization In Thermo Field Dynamics,''
J.\ Phys.\ A \textbf{A22}, 2461 (1989).

%------------------------- NEW RELATIVISTIC NETFD ----------------------------------------


%\cite{Mizutani:2010rk}


\bibitem {mi}Y.~Mizutani and T.~Inagaki,
%``Boltzmann Equation for Relativistic Neutral Scalar Field in Non-equilibrium Thermo Field Dynamics,''
Prog.\ Theor.\ Phys.\ \textbf{125}, 933 (2011) [arXiv:1011.0281 [hep-th]].
%%CITATION = ARXIV:1011.0281;%%


%\cite{Mizutani:2011rt}


\bibitem {miny}Y.~Mizutani, T.~Inagaki, Y.~Nakamura and Y.~Yamanaka,
%``Canonical Quantization for a Relativistic Neutral Scalar Field in Non-equilibrium Thermo Field Dynamics,''
Prog.\ Theor.\ Phys.\ \textbf{126}, , 681 (2011) [arXiv:1105.5952 [hep-th]].
%%CITATION = ARXIV:1105.5952;%%


%\cite{Mizutani:2011jy}


\bibitem {mi2}Y.~Mizutani and T.~Inagaki,
%``Non-Equilibrium Thermo Field Dynamics for Relativistic Complex Scalar and Dirac Fields,''
arXiv:1111.7083 [hep-th].
%%CITATION = ARXIV:1111.7083;%%


%\cite{Green:1987sp}


\bibitem {gsw}M.~B.~Green, J.~H.~Schwarz and E.~Witten,
%``Superstring Theory. Vol. 1: Introduction,''
Cambridge, Uk: Univ. Pr. ( 1987) 469 P. ( Cambridge Monographs On Mathematical Physics)
\end{thebibliography}
\end{document}